\documentclass[10pt]{article}
\usepackage{graphicx}
\usepackage{dcolumn}
\usepackage{tensor}
\usepackage{bm}
\usepackage{float}
\usepackage{color}
\usepackage[usenames,dvipsnames]{xcolor}
\usepackage{listings}
\usepackage{fancyvrb}
\usepackage{slashed}
\usepackage{amsmath}
\usepackage{amssymb}
\usepackage{amsfonts}
\usepackage{enumitem}
\usepackage{listings}
\usepackage{mathtools}
\usepackage{slashed}
\usepackage{verbatim}
\usepackage{hyperref}
\usepackage[margin=1.0in]{geometry}

\newcommand\openone{\leavevmode\hbox{\small1\normalsize\kern-.33em1}}
\def\ee{\end{eqnarray}}

\def\nn{\nonumber}

\newcommand{\<}{\langle}
\renewcommand{\>}{\rangle}

\newcommand{\be}{\begin{eqnarray}}
\newcommand{\en}{\end{eqnarray}}
\newcommand{\bea}[1]{\left(\begin{array}{#1}}
\newcommand{\ena}{\end{array}\right)}
\newcommand{\ba}{\begin{eqnarray}}
\newcommand{\ea}{\end{eqnarray}}
\newcommand{\CL}{\mathcal{L}}
\newcommand{\CM}{\mathcal{M}}
\newcommand{\CO}{\mathcal{O}}
\newcommand{\CH}{\mathcal{H}}

\newcommand\lrnab{\raise .8ex\hbox{$^\leftrightarrow$} \hspace{-8.8pt}
\nabla}
\newcommand\lnab{\raise .8ex\hbox{$^\leftarrow$} \hspace{-9.8pt}
\nabla}
\newcommand\rnab{\raise .8ex\hbox{$^\rightarrow$} \hspace{-9.8pt}
\nabla}

\begin{document}
\title{Model-independent  WIMP Scattering Responses and Event Rates:\\
A Mathematica Package for Experimental Analysis}

\author{Nikhil Anand$^1$, A.~Liam Fitzpatrick$^2$,  W. C. Haxton${^1}$}
\maketitle

\begin{center}
{\it $^1$ Dept. of Physics, University of California, Berkeley, 94720,
and Lawrence Berkeley National Laboratory} \\
{\it $^2$ Stanford Institute for Theoretical Physics, Stanford University, Stanford, CA 94305} \\
\end{center}

\begin{abstract} 
A model independent formulation of WIMP-nucleon scattering was recently developed in Galilean-invariant
effective field theory and embedded in the nucleus, determining the most general WIMP-nucleus elastic response. This formulation shows that the standard description of WIMP elastic scattering in terms spin-dependent and
spin-independent responses frequently fails to identify the dominant operators governing
the scattering, omitting
four of the six responses allowed by basic symmetry considerations.  Consequently comparisons made
between experiments that are based on a spin-independent/spin-dependent analysis can be misleading
for many candidate interactions,
mischaracterizing the magnitude and multipolarity (e.g., scalar or
vector) of the scattering.  The new responses are associated with velocity-dependent WIMP
couplings and correspond to
familiar electroweak nuclear operators such as the
orbital angular momentum $\vec{l}(i)$ and the spin-orbit interaction $\vec{\sigma}(i) \cdot \vec{l}(i)$.  Such
operators have distinct selection rules and coherence properties, and thus open up new opportunities for using low-energy
measurements to constrain ultraviolet theories of dark matter.

The community's reliance on simplified descriptions of WIMP-nucleus interactions reflects the
absence of analysis tools that integrate general theories of dark matter with standard treatments
of nuclear response functions.  To bridge this gap, we have constructed a public-domain Mathematica package for WIMP analyses based on our effective theory formulation.  Script inputs are
1) the coefficients of the effective theory, through which one can characterize
the low-energy consequences of arbitrary ultraviolet theories of WIMP interactions; and 2) one-body density matrices for 
commonly used targets, the most compact description
of the relevant nuclear physics.  The generality of the effective theory expansion guarantees
that the script will remain relevant as new ultraviolet theories are explored; the use of
density matrices to factor the nuclear physics from the particle physics will allow nuclear structure
theorists to update the script as new calculations become available, independent of 
specific particle-physics contexts.  The Mathematica
package outputs the resulting response functions (and associated form factors) and also 
the differential event rate, once a galactic WIMP velocity profile is specified, and thus in its
present form provides a complete framework for experimental analysis.  The Mathematica script requires
no \textit{a priori} knowledge of the details of the non-relativistic effective field theory or nuclear physics, though
the core concepts are reviewed here and in \cite{nonrelEFT}.

\end{abstract}

\section{Introduction\label{sec:intro}} Despite the many successes of the $\Lambda$CDM cosmological model in predicting the macroscopic behavior of dark matter, attempts at an experimentally significant direct detection of the dark matter particle have been unsuccessful and its fundamental nature remains uncertain \cite{2012arXiv1209.3339F} \cite{2005PhR...405..279B}. A promising candidate is a weakly interacting massive particle (`WIMP') that interacts with standard-model particles through a cross section that is suppressed compared to standard electromagnetic interactions. The challenges associated with observing such a particle notwithstanding, experimental techniques are advancing at a rapid pace,
and expectations are high that a definitive measurement of dark matter interactions is imminent.

In ``direct detection" experiments, an important class of dark matter searches, the signals are recoil events following WIMP elastic scattering off target nuclei \cite{PhysRevD.31.3059} \cite{PhysRevD.33.3495} \cite{1996PhR...267..195J}. Many models predict rates for such events consistent with the sensitivities some experiments are now reaching. Most models of WIMPs invoke new physics, such as SUSY or extra dimensions, associated with electroweak symmetry breaking, where new phenomena can appear at scales that, from a particle physics perspective, are quite low, e.g.,  $\gtrsim$ 100 GeV.  However, the momentum transfer in direct detections is still far lower, typically a few hundred MeV or less.
Consequently, effective field theory (EFT) provides a general and very efficient way to characterize experiment
results:  regardless of the complexity or variety of candidate ultraviolet theories
of dark matter, their low-energy consequences can be encoded in a small set of parameters,  such as the mass of the WIMP
and the effective coupling constants describing the strength of the contact coupling of the WIMP to the nucleon or nucleus. 
The information that can be extracted from low-energy experiments can be expressed as constraints 
on the low-energy constants of the EFT.

It has also been conventionally assumed that this momentum transfer is small on the nuclear scale, 
which it is not.  The scattering off the nucleus is treated by modeling the nucleus as a point particle, characterized by a charge and
spin, with the charge and spin couplings sometimes allowed to be isospin dependent.  
This greatly restricts the possible nuclear interactions -- but without justification.  
Recognizing that the point-nucleus approximation is invalid because momentum transfers are generally at least comparable to the inverse nuclear size, attempts have been made to ``repair" the theory by
introducing form factors to account for the finite spatial extent of the nuclear charge and spin densities.
But such a treatment is inadequate and in conflict with standard methods 
for treating related electroweak interactions:  Once momentum transfers reach
that point that $\vec{q} \cdot \vec{x}(i)$, where $\vec{x}(i)$ is the nucleon coordinate within the nucleus, is no longer small, not only form factors, but new operators arise.  These new operators turn out to be
parametrically enhanced for a large class of EFT interactions.

The Galilean-invariant EFT we describe below provides a particularly attractive 
framework for properly treating dark-matter particle scattering. The procedure yields two effective
theories, the first at the level of the WIMP-nucleon scattering amplitude, and the second at the nuclear level, as the embedding of the WIMP-nucleon effective interaction in the nucleus
generates the most general form of the elastic nuclear response.  
Six response functions -- not the two conventionally assumed -- are produced:
\begin{itemize}
\item The new responses typically dominate the elastic cross section for a large class of candidate
interactions involving velocity couplings.  The standard spin independent/spin dependent 
treatment yields amplitudes for such couplings on the order of the WIMP velocity, $\sim 10^{-3}$.
In fact the amplitude is determined by the velocities of bound nucleons, typically $\sim 10^{-1}$.  
\item The neglect of the composite operators not only alters magnitudes, but leads to an 
incorrect dependence of cross sections on basic parameters such as the masses
of the WIMP and target nucleus.
The nuclear physics of the composite nuclear operators is distinctive, with selection rules
unlike those found for the simple charge and spin point operators.  Consequently comparisons made between experiments using
the standard analysis can be quite misleading for a large class of candidate
EFT operators. 
\item Often the standard operators misrepresent even the rank of the response.  
An interaction that in the point nucleus limit appears to be spin-dependent, with amplitude
proportional to matrix elements of $\vec{\sigma}(i)$, may instead produce a much larger scalar
response associated with the composite operator $\vec{\sigma}(i) \cdot \vec{l}(i)$.  Thus a $J=0$ nuclear target may be 
highly sensitive to a given interaction, not blind to it.  
\end{itemize}
None of this physics is exotic: the nuclear physics treatment presented here
is the standard one for semi-leptonic electroweak interactions.  Most of the new
operators that arise in a more careful treatment of the WIMP-nucleus response are familiar
because they are also essential to the correct description of standard-model weak 
and electromagnetic interactions.

The enlarged set of nuclear responses that emerges from a model-independent analysis
has important experimental consequences.   The EFT analysis shows that elastic scattering
can place several new constraints on dark matter properties, in addition to the two 
apparent from the conventional spin-independent/spin-dependent treatment, provided
enough experiments are done.  One can successfully turn the nuclear physics  ``knobs" -- the nuclear responses --
to determine these constraints by utilizing target nuclei with
the requisite ground-state properties. The EFT analysis also shows other ways candidate
interactions can be distinguished, e.g., through the nuclear recoil spectrum (which may depend
on the $v^0$, $v^2$, and $v^4$ moments of the WIMP velocity distribution) or through the dependence
on the mass of the nucleus used in the target.

The basis for our formulation is the description of the WIMP-nucleon interaction in \cite{nonrelEFT} which, building on the work of \cite{fanEFT}, used non-relativistic EFT 
to find the most general low-energy form of that interaction.  The explicit Galilean invariance of the
WIMP-nucleon EFT simplifies the embedding of the resulting effective interaction in 
the nucleus.  This produces a compact and rather elegant form for the WIMP-nucleus elastic cross section
as a product of WIMP and nuclear responses.  The particle physics is isolated in the former.

In \cite{nonrelEFT} the cross section was presented in a largely numerical form,
in principal easy to use but in practice requiring users
to hand-copy lengthy form-factor polynomials.  In contrast, our goals in this paper are to:
1) present the fully general WIMP-nucleus cross section in its most
elegant form, to clarify the physics that can be 
learned from elastic scattering experiments; 2) provide a Mathematica code to evaluate
the expressions, removing the need for either extensive hand copying or a detailed
understanding of operator and matrix element conventions employed in our expressions;
and 3) structure that code to allow easy incorporation of future improved nuclear physics calculations,
so that it will remain useful as the field develops.   We 
believe the script could serve the community as a flexible and very 
adaptable tool for comparing experimental sensitivities and for understanding the
relative significance of experimental limits.

This paper is
organized as follows. We begin in Sec. \ref{sec:sec2} with a brief overview of the EFT construction
of the general WIMP-nucleon Galilean-invariant interaction.  In Sec. \ref{sec:sec3} we describe
the use of this interaction in nuclei.  The EFT scattering probability is shown to consist of six nuclear
response functions, once the constraints of the nearly exact parity and CP of the nuclear ground 
state are imposed.
We point out the differences between our results and spin-independent/spin-dependent formulations,  in order to explicitly demonstrate what physics is lost by assuming a point-nucleus limit.  
In Sec. \ref{sec:sec4} we present differential and total cross sections and rates, discuss integration over the
galactic WIMP velocity profile, and describe cross section scaling properties.  Sec. \ref{sec:sec5} 
we describe the factorization of the operator physics from the nuclear structure that is possible
through the density matrix.  (This will make it possible for nuclear structure theorists to port new
structure calculations into our Mathematica code, without needing to repeat all of the operator
calculations.)   In Sec. \ref{sec:sec6} we construct a
similar interface for particle theorists: we describe the mapping of a very general set of covariant
interactions into EFT coefficients, so that the consequences of a given ultraviolet theory
for WIMP elastic scattering can be easily explored.  In Sec. \ref{sec:sec7} we provide
a tutorial on the code, to help users -- experimentalists interested in analysis, structure
theorists interested in quantifying nuclear uncertainties, or particle theorists interested in 
constraining a candidate ultraviolet theory -- quickly obtain what they need from
the Mathematica script.  Finally in the Appendix, we described some of the algebraic details
one encounters in deriving our master formula for the WIMP-nucleus cross section. As the body of the
paper presents basic results and describes their physical implications, the Appendix is 
intended for those who may be interested in details of the calculations, or possible extensions
of our work.  
The Appendix includes comments on steps in our treatment that are model
dependent or that involve approximations.
We discuss the use of the code for WIMPs with nonstandard properties,
e.g., WIMP-nucleon interactions mediated by  light exchanges.

\section{Effective Field Theory Construction of the Interaction}
\label{sec:sec2}
The idea behind EFT in dark matter scattering is to follow the usual EFT ``recipe", but in a non-relativistic context, by writing down the relevant operators that obey all of the non-relativistic symmetries. In the case of elastic scattering of a heavy WIMP off a nucleon, the Lagrangian density will have the contact form
\be
\mathcal{L}_{\textrm{int}}(\vec{x}) =c ~ \Psi^*_\chi(\vec{x}) \mathcal{O}_\chi \Psi_\chi(\vec{x})~ \Psi^*_N(\vec{x}) \mathcal{O}_N \Psi_N(\vec{x}),
\ee
where the $\Psi(\vec{x})$ are nonrelativistic fields and where the WIMP and
nucleon operators  $\CO_\chi$ and $ \CO_N$ may have vector indices.
The properties of $\CO_\chi$ and $\CO_N$ are then constrained by imposing relevant symmetries. We envision the case where there are a number of candidate interactions
$\CO_i$ formed from the $\CO_\chi$ and  $\CO_N$.
 Working to second order in the momenta, one can construct 
the relevant operators appropriate for use with
Pauli spinors, when constructing the Galilean-invariant amplitude
\be
 \sum_{i=1}^{\cal{N}} \left( c_i^{(n)} \mathcal{O}_i^{(n)} + c_i^{(p)}\mathcal{O}_i^{(p)} \right),
\label{eq:Lag}
\ee
where the coupling coefficients $c_i$ may be different for proton and neutrons.
The number $\cal{N}$ of such operators depends 
on the generality of the particle physics description.  We find that 10 operators arise if we limit our consideration to exchanges
involving up to spin-1 exchanges and to operators that are the
leading-order nonrelativistic analogs of relativistic operators.  Four additional operators
arise if more general mediators are allowed.

This interaction can then be embedded in the nucleus.  The procedure we follow
here -- though we discuss generalizations in the Appendix -- assumes that the nuclear interaction is the 
sum of the WIMP interactions with the individual nucleons in the nucleus.  The nuclear
operators then involve a convolution of the $\CO_i$, whose momenta must now be treated
as local operators appropriate for bound nucleons, with the plane wave associated with
the WIMP scattering, which is an angular and radial operator that can be decomposed 
with standard spherical harmonic methods.  Because momentum transfers are typically
comparable to the inverse nuclear size, it is crucial to carry through such a multipole
decomposition in order to identify the nuclear responses associated with the various
$c_i$s.   The scattering probability is given by the square of the (Galilean) invariant
amplitude $\CM$, a product of WIMP and nuclear matrix elements, averaged over initial
WIMP and nuclear magnetic quantum numbers $M_\chi$ and $M_N$, and summed over final 
magnetic quantum numbers.  The result can be organized
in a way that factorizes the particle and nuclear physics \be
\frac{1}{2j_\chi + 1}\frac{1}{2j_N+1}\sum_{\textrm{spins}} |\mathcal{M}|^2 \equiv \sum_{k} 
\sum_{\tau=0,1} \sum_{\tau^\prime=0,1} R_k \left( \vec{v}_T^{\perp 2}, {\vec{q}^{\,2} \over m_N^2},\left\{c_i^\tau c_j^{\tau^\prime} \right\} \right)
~W_k^{\tau \tau^\prime}( \vec{q}^{\,2} b^2)
\label{eq:RS}
\ee
where the sum extends over products of WIMP response functions $R_k$ and
nuclear response functions $W_k$.   The $R_k$ isolate the particle physics: they depend on
specific combinations of bilinears in the low-energy constants of the EFT -- the 2$\cal{N}$ 
coefficients of Eq. (\ref{eq:Lag}) -- here labeled by isospin $\tau$ (isoscalar, isovector) rather
than the $n,p$ of Eq. (\ref{eq:Lag}) (see below).  The WIMP response functions also depend
on the relative WIMP-target velocity $\vec{v}_T^\perp$, defined below for the nucleon (and 
in Sec. \ref{sec:sec3pt4} for a nucleus), and three-momentum transfer 
$\vec{q}= \vec{p}^{\,\prime}-\vec{p}=\vec{k} - \vec{k}^{\prime}$, where $\vec{p}$ ($\vec{p}^{\,\prime}$)
is the incoming (outgoing) WIMP three-momentum and $\vec{k}$ ($\vec{k}^\prime$) the incoming
(outgoing) nucleon three-momentum.
The  nuclear response functions $W_k$ can be varied by experimentalists, if they explore a variety
of nuclear targets.   
The $W_k$ are functions of $y \equiv (q b/2)^2$, where $b$ is the nuclear size (explicitly the
harmonic oscillator parameter if the nuclear wave functions are expanded in that 
single-particle basis).

EFT provides an attractive framework for analyzing and comparing direct detection experiments.  It simplifies the analysis of WIMP-matter interactions by exploiting an important small parameter: typical velocities of the particles
comprising the dark matter halo are $v/c \sim 10^{-3}$, and thus non-relativistic.  Consequently,
while there may be a semi-infinite number of candidate ultraviolet theories of WIMP-matter interactions,
many of these theories are operationally indistinguishable at low energies.  By organizing the effective field theory in terms of non-relativistic interactions and degrees of freedom, one can significantly simplify the classification of possible operators \cite{nonrelEFT,fanEFT}, while not
sacrificing generality.  In constructing the needed set of independent operators,  the equations of motion are employed to remove redundant operators.  The operators themselves are expressed in terms of quantities that are more directly related to scattering observables at the relevant energy scale, which makes the relationship between operators and the underlying physics more transparent. Furthermore, it becomes trivial to write operators for arbitrary dark matter spin, a task that can be
rather involved in the relativistic case.  

EFT also prevents oversimplification: because it produces a complete
set of effective interactions at low energy, one is guaranteed that the description is general.  Provided
this interaction is then embedded in the nucleus faithfully, it will then produce the most general
nuclear response consistent with the assumed symmetries.  Consequently some very basic questions
that do not appear to be answered in the literature can be immediately addressed.  
How many constraints on dark matter particle interactions can be obtained
from elastic scattering?  
Conversely, what redundancies exist among the EFT's low-energy constants that cannot be resolved,
regardless of the number of elastic-scattering experiments that are done?   

\subsection{Constructing the Nonrelativistic Operators}
Because dark matter-ordinary matter interactions are more commonly described in relativistic
notation, we will begin by considering the nonrelativistic reduction of two familiar 
relativistic interactions.  
We consider the
contact interaction between a spin-1/2 WIMP and nucleon,
\be
\CL_{\rm int}^{\rm SI}(\vec{x}) =c_1~ \bar{\Psi}_\chi(\vec{x}) \Psi_\chi(\vec{x}) ~\bar{\Psi}_N(\vec{x}) \Psi_N(\vec{x}) .
\label{eq:scalar}
\ee
We employ Bjorken and Drell gamma matrix conventions including their spinor normalization
(1 instead of the $2m$
used in \cite{nonrelEFT}).  Because of the change, the $c$s defined here, which carry dimensions of 1/mass$^2$,
differ from those of Ref. \cite{nonrelEFT}.  The relativistic fields of Eq. (\ref{eq:scalar})
include spinors $U_\chi(p)$ and $U_N(p)$ that can be written at low momenta as
\be
U_\chi (p) =\sqrt{{E+m \over 2m}} \left( \begin{array}{cc}  \xi_\chi \\[0.2cm] \displaystyle\frac{\vec{\sigma}\cdot \vec{p}}{E+m_\chi}  \xi_\chi \end{array} \right) \sim   \left( \begin{array}{cc}  \xi_\chi \\[0.2cm] \displaystyle\frac{\vec{\sigma}\cdot \vec{p}}{2m_\chi}  \xi_\chi \end{array} \right).
\ee
and consequently to leading order in $p/m_\chi$ and $p/m_N$, we obtain the nonrelativistic operator
\be
c_1 ~ 1_\chi 1_N~\equiv ~c_1~ \CO_1
\label{eq:spini}
\ee
that would be evaluated between Pauli spinors $\xi_\chi$ and $\xi_N$, to form the nonrelativistic
analog of the invariant amplitude.  Here $\CO_1$ is one of the EFT
operators we introduce below. The non-relativistic form of another interaction
\be
\CL_{\rm int}^{\rm SD} =c_4~ \bar{\chi} \gamma^\mu \gamma^5 \chi \bar{N} \gamma_\mu \gamma^5 N.
\label{eq:axialvector}
\ee
 is also easily taken.  In this case, the dominant contribution in the non-relativistic limit comes from the spatial indices, with  $\bar{\chi} \gamma^i \gamma^5 \chi \sim \xi_\chi^\dagger \sigma^i \xi_\chi$. The $\sigma^i$ matrix here is just twice the particle spin $S^i$, so we obtain the
nonrelativistic operator
\be
 -4c_4~ \vec{S}_\chi \cdot \vec{S}_N~\equiv~~-4c_4~ \CO_4.
\label{eq:spind}
\ee
Equations (\ref{eq:spini}) and (\ref{eq:spind}) correspond to the spin-independent and
spin-dependent operators used so frequently in experimental analyses.

One could continue in this manner, constructing all possible relativistic interactions, and considering their 
nonrelativistic reductions.  But this is unnecessary.  One advantage of non-relativistic EFT is its systematic treatment of interactions, including those with momentum-dependence.  Operators can be constructed not only with the three-vectors $\vec{S}_\chi$ and $\vec{S}_N$, but also using the momenta of the WIMP and nucleon. Of the four momenta involved in the scattering (two incoming and two outgoing), only two combinations are physically relevant due to inertial frame-independence and momentum conservation.  It is convenient to work with the frame-invariant quantities, the momentum transfer $\vec{q}$ and the WIMP-nucleon
relative velocity,
\be
\vec{v} \equiv \vec{v}_{\chi, \rm in} - \vec{v}_{N, \rm in}.
\ee
It is also useful to construct the related quantity
\be
\vec{v}^\perp = \vec{v} + \frac{\vec{q}}{2\mu_N} = {1 \over 2} \left( \vec{v}_{\chi,\mathrm{in}}+\vec{v}_{\chi,\mathrm{out}} - \vec{v}_{N,\mathrm{in}} - \vec{v}_{N,\mathrm{out}} \right) =
 {1 \over 2} \left( {\vec{p} \over m_\chi}+{\vec{p}^{\, \prime} \over m_\chi} - {\vec{k} \over m_N} - {\vec{k}^{\, \prime} \over m_N} \right)
\label{eq:vperp}
\ee
which satisfies $\vec{v}^\perp \cdot \vec{q} =0$ as a consequence of energy conservation.  Here $\mu_N$ is the WIMP-nucleon reduced mass.  It was shown in \cite{nonrelEFT} that operators are guaranteed to be Hermitian if they are built out of the following four  three-vectors,
 \ba
 && i {\vec{q} \over m_N}, \ \ \ \ \vec{v}^\perp, \ \ \ \ \vec{S}_\chi, \ \ \ \ \vec{S}_N.
 \ea
Here (in another departure from \cite{nonrelEFT}) we have introduced $m_N$ as an convenient scale to render $\vec{q}/m_N$ and the constructed
$\CO_i$ dimensionless: the choice of this scale is not arbitrary, as it leads to an EFT power counting
in nuclei that is particularly simple, as we discuss in Sec. \ref{sec:sec2pt3} and in greater 
detail in Sec. \ref{sec:sec4pt3}.
The relevant interactions that we can construct from these three-vectors and that can be 
associated with interactions involving only spin-0 or spin-1 mediators are
\be
\CO_1 &=& 1_\chi 1_N \nn \\
\CO_2 &=& (v^\perp)^2 \nn\\
 \CO_3 &=& i \vec{S}_N \cdot ({\vec{q} \over m_N} \times \vec{v}^\perp) \nn\\
\CO_4 &=& \vec{S}_\chi \cdot \vec{S}_N \nn\\
\CO_5 &=& i \vec{S}_\chi \cdot ({\vec{q} \over m_N} \times \vec{v}^\perp)  \nn\\
 \CO_6&=& (\vec{S}_\chi \cdot {\vec{q} \over m_N}) (\vec{S}_N \cdot {\vec{q} \over m_N}) \nn \\
\CO_7 &=& \vec{S}_N \cdot \vec{v}^\perp \nn\\
\CO_8 &=& \vec{S}_\chi \cdot \vec{v}^\perp \nn\\ 
 \CO_9 &=& i \vec{S}_\chi \cdot (\vec{S}_N \times {\vec{q} \over m_N}) \nn\\
\CO_{10} &=& i \vec{S}_N \cdot {\vec{q} \over m_N}  \nn\\
\CO_{11} &=& i \vec{S}_\chi \cdot {\vec{q} \over m_N}
\label{eq:ops}
\ee
These 11 operators were discussed in \cite{nonrelEFT}.  We retain 10 these here, discarding
$\CO_2$, as this operator cannot  be obtained from the leading-order non-relativistic reduction of a manifestly relativistic operator (see Sec. \ref{sec:sec6}).  
$\CO_2$
was retained in \cite{nonrelEFT} because it corrects a coherent operator.  
However as similar operators arise as corrections to other interactions --
e.g., the reduction of an axial vector-axial vector interaction generates $\CO_2^b \equiv \vec{v}^\perp \cdot
\vec{S}_\chi \vec{v}^\perp \cdot \vec{S}_N$ as well as  $\CO_4$ --
here we take the view
that it is more consistent to retain only those operators found in Table \ref{table:LWL}
of Sec. \ref{sec:sec6}.

We classify these operators as  LO, NLO, and N$^2$LO, depending on the total number of
momenta and velocities they contain.  We will see in Sec. \ref{sec:sec4pt3} that these designations
correspond to total cross sections that scale as $v^0$, $v^2$, or $v^4$, where $v$ is the WIMP
velocity in the laboratory frame.

In addition, one can construct the following operators that do not arise for traditional spin-0 or spin-1 mediators
\be
\CO_{12} &=& \vec{S}_\chi \cdot (\vec{S}_N \times \vec{v}^\perp) \nn\\ 
\CO_{13} &=&i (\vec{S}_\chi \cdot \vec{v}^\perp  ) (  \vec{S}_N \cdot {\vec{q} \over m_N})  \nn \\
\CO_{14} &=& i ( \vec{S}_\chi \cdot {\vec{q} \over m_N})(  \vec{S}_N \cdot \vec{v}^\perp )  \nn\\
\CO_{15} &=& - ( \vec{S}_\chi \cdot {\vec{q} \over m_N}) ((\vec{S}_N \times \vec{v}^\perp) \cdot {\vec{q} \over m_N})  \nn\\
\CO_{16} &=& - ((\vec{S}_\chi \times \vec{v}^\perp) \cdot {\vec{q} \over m_N})( \vec{S}_N \cdot {\vec{q} \over m_N}) .
\ee
It is easy to see that $\CO_{16}$ is linearly dependent on $\CO_{12}$ and $\CO_{15}$,
\be
\CO_{16} &=& \CO_{15} + {\vec{q}^{\,2} \over m_N^2}  \CO_{12} ,
\ee
and so should be eliminated.  Operator $\CO_{15}$ is
cubic in velocities and momenta, generating a total cross section of order $v^6$ (N$^3$LO).
It is retained because it arises as the leading-order nonrelativistic limit of certain covariant 
interactions constructed in Sec. \ref{sec:sec6}. 

Each operator
can have distinct couplings to protons and neutrons.  Thus the EFT interaction 
we employ in this paper takes the form
\be
 \sum_{\alpha=n,p} ~\sum_{i=1}^{15} c_i^{\alpha} \CO_i^{\alpha},~~~~c_2^{\alpha} \equiv 0.
\label{eq:int}
\ee
 One can factorize 
the space-spin and proton/neutron components of Eq. (\ref{eq:int}) by introducing isospin, which
is also useful as an approximate symmetry of the nuclear wave functions.  Thus
an equivalent form for our interaction is
\be
 \sum_{i=1}^{15} ( c_i^{0} 1 + c_i^{1} \tau_3) \CO_i = \sum_{\tau=0,1} \sum_{i=1}^{15} c_i^\tau \CO_i t^\tau, ~~~~c_2^{\mathrm{0}}=c_2^{\mathrm{1}} \equiv 0,
\label{eq:Lagrangian}
\ee
where the isospin state vectors, operators, and couplings are
\be
|p\rangle = \left( \begin{array}{c} 1 \\ 0 \end{array} \right) ~~~|n\rangle = \left( \begin{array}{c} 0 \\ 1 \end{array} \right)~~~1 \equiv \left( \begin{array}{cc} 1 & 0 \\ 0 & 1 \end{array} \right) ~~~\tau_3 \equiv \left( \begin{array}{cc} 1 & 0 \\ 0 & -1 \end{array} \right)~~~c_i^{0} = {1 \over 2} (c_i^{\mathrm{p}}+c_i^{n})~~~c_i^{1} = {1 \over 2} (c_i^{p}-c_i^{n})
\ee
and where the isospin operators are defined by
\be
t^0 \equiv 1~~~t^1 \equiv \tau_3.
\label{eq:iso}
\ee

The EFT has a total of 28 parameters, associated with 14 space/spin operators each
of which can have distinct couplings to protons and neutrons.  If we exclude operators that are not
associated with spin-0 or spin-1 mediators, 10 space/spin operators and 20 couplings remain.

\subsection{Units: Inputing the $c_i$s into the Mathematica Script}
\label{sec:sec2pt2}
The interactions of Eqs. (\ref{eq:scalar}) and (\ref{eq:axialvector}) are very similar to familiar vector-vector and
axial vector-axial vector interactions of the standard model.  For example, the replacement
\be
c_4 \CO_4 t^1 \equiv c_4 \CO_4 \tau_3 \rightarrow {G_F \over \sqrt{2}} \CO_4 \tau_{\pm} 
\ee
where $G_F \sim 1.166 \times 10^{-5} \mathrm{~GeV}^{-2}$ is the Fermi constant and $\tau_{\pm}$ is the isospin raising or lowering operator, yields the Gamow-Teller interaction familiar in
low-energy charged-current neutrino scattering off nuclei.  $G_F$ defines a standard-model
weak interaction mass scale
\be
m_v \equiv \langle v \rangle = (2 G_F)^{-1/2} = 246.2 \mathrm{~GeV}
\ee
where $\langle v \rangle$ is the Higgs vacuum expectation value.  

Much of the theoretical motivation for WIMP searches is connected with the ``WIMP miracle,"
that weakly interacting massive particles will naturally freeze out in the early universe, when
their annihilation rate falls behind the expansion rate, to produce a relic density today
consistent with the dark matter density.  The experimental program is focused on probing
at and beyond the weak scale for dark matter interactions.
It is a natural scale, then, for characterizing the strengths of
interactions now being constrained by experiments. Consequently, in our
Mathematica script all of the $c_i$s are input in weak-interaction units, defined as
\be
\mathrm{input~}c_i = 1 \Rightarrow c_i = 1/m_v^2
\ee
Thus an input of  $c_i=10,~1,\mathrm{~and}~0.1$
  converts to $c_i=10/m_v^2,~1/m_v^2,
\mathrm{~and~} 0.1/m_v^2$,
producing interactions of strength 10, 1, and 1/10th of weak, and cross sections 100, 1, and 1/100th of weak,
respectively.

\subsection{EFT Power Counting and $\vec{q}/m_N$: Parametric Enhancement}
\label{sec:sec2pt3}
The EFT formulation leads to a attractive power counting that is helpful in understanding
the dependence of laboratory total cross sections on the physically relevant parameters - the WIMP
velocity $\vec{v}$, the ratio of the WIMP-nuclear target reduced mass $\mu_T$ to $m_N$, and the ratio of $\mu_T$ to the inverse nuclear size.   The scaling rules we will discuss in Sec. \ref{sec:sec4pt3}
take on a simple form if $m_N$ is used to construct the dimensionless quantity $\vec{q}/m_N$,
a parameter related to the relative velocities of nucleons bound in the nucleus, as explained below.  
The fact that this velocity
is much greater than the WIMP velocity leads to a parametric enhancement of the certain
``composite operator" contributions to cross sections.

The introduction of the scale $m_N$ would be arbitrary if
we limit ourselves to WIMP-nucleon scattering.  Any other choice 
would simply lead to the same scaling of the total cross section on $\mu_T/m_N$, but with the $m_N$ in the denominator replaced by that new scale.  There is a single relative velocity $\vec{v}^\perp_T$ in the 
WIMP-nucleon system, associated with the Jacobi coordinate, the distance between
the WIMP and the nucleon.

But in a system consisting of a WIMP
and a nucleus containing $A$ nucleons, there are $A$ independent Jacobi coordinates, and
$A$ associated independent velocities.  
Any WIMP-nucleon velocity-dependent interaction summed over the
nucleons in a nucleus must of course involve all of these velocities.  One of these can be
chosen to be the WIMP-target relative velocity, measured with respect to the center-of-mass
of the nucleus, or $\vec{v}^\perp_T$, the  analog of the single WIMP-nucleon velocity.
But in addition to this velocity, there are $A-1$ others
associated with the $A-1$ independent Jacobi inter-nucleon coordinates.  These velocities
are Galilean invariant intrinsic nuclear operators.  

An internal velocity carries negative parity, and thus its nuclear matrix element vanishes due to 
the nearly exact parity of the nuclear ground state.  
However, because the nucleus is composite, the nuclear operators built from $\CO_i$ are accompanied by 
an additional spatial operator $e^{-i \vec{q} \cdot \vec{x}(i)}$.  A threshold operator carrying the
requisite positive parity can thus be formed by combining $i \vec{q} \cdot \vec{x}(i)$ with 
$\vec{v}(i) = \vec{p}(i)/m_N$.   But $\vec{p}(i)$ and $\vec{x}(i)$ are conjugate operators: the
larger the nuclear size, the smaller the nucleon momentum scale.  Thus when $\vec{p}(i)$ and
$\vec{x}(i)$ are combined to form interactions, one obtains operators such as $ \vec{l}(i)$,
the orbital angular momentum, that have no
associated scale: the single-particle eigenvalues of $l_z(i)$ are integers.  (Operators built from 
such internal nuclear coordinates will be called composite operators.)   Thus scattering
associated with internal velocities is governed by the parameters multiplying $\vec{p}(i)$
and $\vec{x}(i)$, which form the dimensionless ratio $\vec{q}/m_N$.  This dimensionless parameter
emerges directly from the physics -- it is not put in by hand.

Thus we see that  $\vec{q}/m_N$ is associated with the typical velocity
of bound nucleons, $\sim 1/10$.  The composite operators constructed from nucleon velocities are
enhanced relative to those associated with $\vec{v}^\perp_T$ by the ratio of 
$\vec{q}/m_N$ to $\vec{v}^\perp_T$, or
$\sim 100$.   The standard point-nucleus treatment of WIMP scattering retains
only the effects of $\vec{v}^\perp_T$.  We will find in Sec. \ref{sec:sec4pt3} that the enhancement
associated with $\vec{q}/m_N$ leads to an 
increased sensitive to derivative couplings of $\sim 10 (\mu_T/m_N)^2$ in the total cross section,
relative to point nucleus treatments.

\section{The Nuclear Response in EFT \label{sec:sec3}}
Cross sections or rates for WIMP-nucleon/nucleus scattering can be expressed as simple kinematic 
integrals over a fundamental particle-nuclear function, 
the square of the invariant amplitude averaged over initial WIMP and nuclear spins and
summed over final spins.  The key result of this subsection is the calculation of this quantity
for the EFT interaction.

Because much of the literature employs analyses based on the spin-independent/spin-dependent formulation, we begin by
considering two limits in which such a result is obtained.  
One way to obtain a spin-independent/spin-dependent result while still using a very general
interaction, such as the EFT form developed here, is to treat the nucleus as a point particle.  
Effectively one replaces $e^{-i \vec{q} \cdot \vec{x}(i)} \CO_i$ by $\CO_i$, despite the fact
that $\vec{q} \cdot \vec{x}(i)$ is typically $\sim$ 1.  Alternatively, one can simply restrict
 the operators initially to $\CO_1$ and $\CO_4$, the two LO operators in our EFT list.  Then one can proceed
to do a full nuclear calculation, including form factors.   Unfortunately many reasonably
candidate dark-matter interactions do not have the $\CO_1/\CO_4$ form, and thus cannot be represented
in this way.

The spin-independent/spin-dependent results given below can be compared with the results from 
the model-independent formulation presented in Sec. \ref{sec:sec3pt4}.  This is the simplest way to illustrate
what physics is lost in a spin-independent/spin-dependent formulation.  

\noindent
\subsection{The nucleon calculation}  One could in principle detect WIMPs through their 
elastic scattering off free protons and (hypothetically) neutrons.  Such a target can be
treated as a point because the inverse nucleon size is large compared to typical momentum
transfers in WIMP scattering.    In this case the EFT Galilean-invariant amplitude corresponding
to Eq. (\ref{eq:Lagrangian}) for a proton target becomes
\begin{eqnarray}
\CM &=& \langle \vec{p}^{\,\prime} S_\chi m_\chi;~\vec{k}^\prime S_N={1 \over 2}~ m_N~ T_N={1 \over 2} m_T={1 \over 2} | ~\CH~ |\vec{p} S_\chi m_\chi;~\vec{k}  S_N={1 \over 2}~ m_N~T_N={1 \over 2} m_T={1 \over 2} \rangle
\end{eqnarray}
where we have introduced the proton's isospin quantum numbers for consistency
with the isospin form of our Hamiltonian, Eq. (\ref{eq:Lagrangian}).
An elementary calculation then yields the square of the invariant amplitude, averaged over initial
spins and summed over final spins, for WIMP scattering off a proton
\begin{eqnarray}
\frac{1}{2j_\chi + 1}~\frac{1}{2}\sum_{\textrm{spins}} |\mathcal{M}|^2_\mathrm{proton} =\left[ c_1^{p \,2} + {j_\chi(j_\chi+1) \over 3}  \left( {\vec{q}^{\, 2} \over m_N^2} \vec{v}_T^{\perp 2} c_5^{p \, 2}
+ \vec{v}_T^{\perp 2} c_8^{p \, 2} + {\vec{q}^{~2} \over m_N^2}  c_{11}^{p \,2} \right) \right] |M_{F;p}|^2 ~~~~~~~\nonumber \\
~~~+{1 \over 12} \left[ \left( {\vec{q}^{\, 2} \over m_N^2} \vec{v}_T^{\perp 2} c_3^{p \, 2}  + \vec{v}_T^{\perp 2} c_7^{p \, 2} +  {\vec{q}^{\, 2} \over m_N^2}  c_{10}^{p \, 2} \right)
+{j_\chi(j_\chi+1) \over 3}  \left( 3 c_4^{p \, 2} +  2 {\vec{q}^{\, 2} \over m_N^2} ( c_4^p c_6^p+c_9^{p \,2}) +  {\vec{q}^{\, 4} \over m_N^4} c_6^{p \,2}+~~  \right. \right.  \nonumber \\
\left. \left. ~~ 
+ 2 \vec{v}_T^{\perp 2}  c_{12}^{p \, 2} + {\vec{q}^{\,2} \over m_N^2} \vec{v}_T^{\perp \,2} (c_{13}^{p \, 2} + c_{14}^{p \, 2}-2 c_{12}^p c_{15}^{p}) + {\vec{q}^{\,4} \over m_N^4} \vec{v}_T^{\perp 2} c_{15}^{p \,2} \right)  \right] |M_{GT;p}|^2~~~~~~
\label{eq:nucleon}
\end{eqnarray}
where $||$ denotes a matrix element reduced in spin.   The spin-independent
(or Fermi) and spin-dependent (or Gamow-Teller) operators evaluated between nonrelativistic Pauli
spinors have the values
\be
|M_{F;p}|^2 \equiv{1 \over 2}  |\langle 1/2 || 1 || 1/2 \rangle |^2 =1 ~~~~~~~~|M_{GT;p}|^2 \equiv {1 \over 2}  |\langle 1/2 || \sigma || 1/2 \rangle |^2=3
\ee
where the subscript $p$ is an explicit reminder that this is a proton matrix element.
If this result is integrated over phase space, one obtains a cross section that depends the
two particle-physics quantities within the square brackets of Eq. (\ref{eq:nucleon}), with the
associated kinematic factors evaluated by averaging over the WIMP velocity distribution.

\noindent
\subsection{The spin-independent/spin-dependent nuclear form: Point nucleus limit}
A charge-independent/charge-dependent transition probability
\be
\frac{1}{2j_\chi + 1}~\frac{1}{2j_N+1}\sum_{\textrm{spins}} |\mathcal{M}|^2_\mathrm{pt~nucleus} 
\ee
for a point nucleus of spin $j_N$ is obtained by making two substitutions in
Eq. (\ref{eq:nucleon}).  First, the proton Fermi and Gamow-Teller matrix elements are replaced by their nuclear analogs
\begin{eqnarray}
|M_{F;p}|^2& \rightarrow& |M_{F;p}^N(0)|^2 \equiv \left[{1 \over 2j_N+1}  |\langle j_N || \sum_{i=1}^A {1 + \tau_3(i) \over 2} || j_N \rangle |^2 \right]=Z^2 \nonumber \\
|M_{GT;p}|^2 &\rightarrow& |M_{GT;p}^N(0)|^2 \equiv  \left[{1 \over 2j_N+1}  |\langle j_N|| \sum_{i=1}^A {1 + \tau_3(i) \over 2}\sigma(i) || j_N \rangle |^2 \right] 
\label{eq:point}
\end{eqnarray}
where we have assumed that the WIMP coupling is only to protons --
enforced by the introduction of the isospin operators --
to produce a result analogous to Eq. (\ref{eq:nucleon}).   Second, the velocity $\vec{v}_T^\perp$ that
in the nucleon case represented the WIMP-nucleon relative velocity now becomes the
analogous parameter measured with respect to the nuclear center of mass.  There are no
intrinsic nuclear velocities because the nucleus is a point.

The resulting expression has two defects, the absence of nuclear form factors that are required because
the momentum transfer is significant on nuclear scales; and the absence
of contributions due to the intrinsic nucleon velocities, which generate the operators most sensitive to
several of our EFT interactions $\CO_i$.

\noindent
\subsection{The spin-independent/spin-dependent nuclear form: Allowed limit} 
The spin-independent/spin-dependent result most often seen in the literature properly
accounts for the momentum transfer in the scattering, but simplifies the WIMP-nucleon
operator by assuming it is formed from a linear
combination of $\CO_1$ and $\CO_4$, despite any evidence to support such an assumption.

The WIMP-nucleus interaction is written as the sum over WIMP interactions with the
bound nucleons, deriving from $\CO_1$ and $\CO_2$  the WIMP interactions with
the respective extended nuclear charge and spin-current densities
\begin{eqnarray}
 1_\chi  \rho_N(\vec{x}) = 1_\chi~ \sum_{i=1}^A  (c_1^0 + c_1^1 \tau_3(i)) e^{-i \vec{q} \cdot \vec{x}_i} \rightarrow  c_1^{p}~1_\chi~ \sum_{i=1}^A  {1 +  \tau_3(i) \over 2} e^{-i \vec{q} \cdot \vec{x}_i} ~~~~~~~~~~~~~~\nonumber \\
\vec{S}_\chi \cdot \vec{j}_N(\vec{x})= \vec{S}_\chi \cdot \sum_{i=1}^A  (c_4^0 +c_4^1 \tau_3(i)) {\vec{\sigma}(i) \over 2}  e^{-i \vec{q} \cdot \vec{x}_i} \rightarrow c_4^p~  \vec{S}_\chi \cdot  \sum_{i=1}^A  {1 + \tau_3(i) \over 2} {\vec{\sigma}(i) \over 2}  e^{-i \vec{q} \cdot \vec{x}_i}
\label{eq:simpledensities}
\end{eqnarray}
where on the right we have again
simplified the result by restricting the couplings to protons, to allow comparisons with Eqs. (\ref{eq:nucleon}) and (\ref{eq:point}).
 
The spin averaged/summed transition probability can be easily evaluated by the spherical harmonic methods 
outlined in the Appendix, yielding
\begin{eqnarray}
\frac{1}{2j_\chi + 1}~\frac{1}{2j_N+1}\sum_{\textrm{spins}} |\mathcal{M}|^2 &=&
c_1^{p \,2} \left[ {4 \pi \over 2 j_N+1}   \sum \limits_{J=0,2,...}^\infty |\langle j_N || \sum \limits_{i=1}^A M_J(q x_i) {1 + \tau_3(i) \over 2} || j_N \rangle |^2  \right] ~~~~~~~\nonumber \\
&+& c_4^{p \,2}~ {j_\chi (j_\chi+1) \over 12} \left[ {4 \pi \over 2 j_N+1} 
\sum \limits_{J=1,3,...}^\infty \left( |\langle j_N ||~ \sum \limits_{i=1}^A \Sigma^{\prime \prime}_J(q x_i) {1 + \tau_3(i) \over 2} || j_N \rangle |^2 \right. \right.  \nonumber \\
&&\left. \left. ~~~~~~~~~~~~~~~~~~~~~~~~~~~~~~~~~~+ |\langle j_N || \sum \limits_{i=1}^A \Sigma^{\prime}_J(q x_i) {1 + \tau_3(i) \over 2} || j_N \rangle |^2 \right) \right] \nonumber \\
&\equiv& c_1^{p \,2} |M^N_{F;p}(0)|^2 F_F^{p \,2}(q^2) 
+ c_4^{p \,2}~ {j_\chi (j_\chi+1) \over 12}~ |M^N_{GT;p}(0)|^2  F_{GT}^{p \,2}(q^2) 
\label{eq:allowed}
\end{eqnarray}
Here $M_J(q x_i)$ is the charge multipole operator and $\Sigma_J^{\prime \prime}(q x_i)$ and $\Sigma_J^{\prime}(q x_i)$ are the longitudinal
and transverse spin multipole operators of rank $J$, which are standard in treatments of 
electroweak nuclear interactions, and will be defined below.  The
assumption of nuclear wave functions of good parity and CP restricts the sums to even and
odd $J$, respectively. 

The form factors $F_F^p(q^2)$ and $F_{GT}^p(q^2)$ are defined so that $F_F^p(0)=F_{GT}^p(0) = 1$,
and can be computed from a nuclear model
\begin{eqnarray}
F_F^{p \,2}(q^2) &=&\frac{  \sum \limits_{J=0,2,...}^\infty |\langle j_N || \sum \limits_{i=1}^A M_J(q x_i) {1 + \tau_3(i) \over 2} || j_N \rangle |^2 }{ {1 \over 4 \pi} |\langle j_N|| \sum \limits_{i=1}^A {1 + \tau_3(i) \over 2} || j_N \rangle |^2} \nonumber \\
F_{GT}^{p \,2}(q^2) &=&\frac { \sum \limits_{J=1,3,...}^\infty \left( |\langle j_N || \sum \limits_{i=1}^A \Sigma^{\prime \prime}_J(q x_i) {1 + \tau_3(i) \over 2} || j_N \rangle |^2+ |\langle j_N || \sum \limits_{i=1}^A \Sigma^{\prime}_J(q x_i) {1 + \tau_3(i) \over 2} || j_N \rangle |^2 \right) }{ {1 \over 4 \pi} |\langle j_N || \sum \limits_{i=1}^A {1 + \tau_3(i) \over 2} \sigma(i) || j_N\rangle |^2 }. \nonumber \\
~
\label{eq:formfactor}
\end{eqnarray}
The spin form factor has the above form because of the identity
\be
 \vec{S}_\chi \cdot \vec{S}_N \equiv (\vec{S}_\chi \cdot \hat{q} ) (\vec{S}_N \cdot \hat{q}) +  (\vec{S}_\chi \times \hat{q} )\cdot  (\vec{S}_N \times \hat{q})  
 \label{eq:equal}
 \ee
 where $\hat{q}$ is the unit vector along the momentum transfer to the nucleus.   Thus the
 use of $\CO_4$ implies equal couplings to the longitudinal and transverse spin operators
 $\Sigma_J^{\prime \prime}$ and $\Sigma_J^\prime$, which cannot interfere if one sums
 over spins.  In a more general treatment of the WIMP-nucleon interaction, these
 operators would be independent.  For example, 
 in the EFT expansion $\CO_4=\vec{S}_\chi \cdot \vec{S}_N$ and
 $\CO_6=(\vec{S}_\chi \cdot \vec{q} ) (\vec{S}_N \cdot \vec{q})$ have distinct coefficients.

Often in the literature $F_F^p(q^2)$ and $F_{GT}^p(q^2)$ are not calculated microscopically,
but are represented by simple phenomenological forms.  

The operators $M_J$, $\Sigma_J^{\prime \prime}$, and $\Sigma_J^\prime$ are the 
vector charge, axial longitudinal, and axial transverse electric multipole operators familiar
from electroweak nuclear physics.  The latter two operators are also 
frequently designated as $L_J^5$
and $T_J^{\mathrm{el} \, 5}$ in the literature, to emphasize their multipole and axial character.  

While we have simplified the above expressions by assuming all couplings are to
protons, to allow a comparison with our free-proton result, the expressions for arbitrary
isospin are also simple 
\begin{eqnarray}
\frac{1}{2j_\chi + 1}~\frac{1}{2j+1}\sum_{\textrm{spins}} |\mathcal{M}|^2 &=&
 {4 \pi \over 2  j_N+1} \left[ \sum \limits_{J=0,2,...}^\infty |\langle j_N || \sum \limits_{i=1}^A M_J(q x_i) \left(c_1^0 + c_1^1 \tau_3(i)\right) || j_N \rangle; |^2  \right. ~~~~~~~~~~~~~~ \nonumber \\
&+& {j_\chi (j_\chi+1) \over 12} \sum \limits_{J=1,3,...}^\infty \left( |\langle j_N || \sum \limits_{i=1}^A \Sigma^{\prime \prime}_J(q x_i) \left(c_4^0 + c_4^1 \tau_3(i) \right) || j_N \rangle |^2 \right.\nonumber \\
 &&~~~~~~~~~~~~~~~~~~~~+ \left. \left.  |\langle j_N || \sum \limits_{i=1}^A \Sigma^{\prime \prime}_J(q x_i) \left(c_4^0 +c_4^1  \tau_3(i) \right) || j_N \rangle |^2 \right) \right] 
\end{eqnarray}

\noindent
\subsection{The general EFT form of the WIMP-nucleus response}
\label{sec:sec3pt4}
The general form of the WIMP-nucleus interaction consistent with the assumption of
nuclear ground states with good P and CP can be derived by building an EFT at the
nuclear level, or by embedding the EFT WIMP-nucleon interaction into the nucleus,
without making assumptions of the sort just discussed.  
 We follow the second strategy here, as it allows us to connect
the nuclear responses back to the single-nucleon interaction and consequently to
the ultraviolet theories which map onto that single-nucleon interaction, on nonrelativistic reduction.

While the calculation is not difficult, we relegate most of the details to the Appendix,
giving just the essentials here.  First, the basic model assumption is that the nuclear
interaction is the sum of the interactions of the WIMP with the individual nucleons in the nucleus.  Thus
the mapping from the nucleon-level effective operators to nuclear operators is made
by the following generalization of Eq. (\ref{eq:Lagrangian}),
\be
\sum_{\tau=0,1} \sum_{i=1}^{15} c_i^\tau \CO_i t^\tau ~\rightarrow~ \sum_{\tau=0,1} \sum_{i=1}^{15} c_i^\tau \sum_{j=1}^A \CO_i(j) t^\tau(j), ~~~c_2^0=c_2^1=0.
\ee
Now the nuclear operators appearing in this expression are built from $i \vec{q}/m_N$, 
a c-number, $\vec{S}_N$, which acts on intrinsic nuclear coordinates, and 
the relative velocity operator $\vec{v}^\perp$,
which now represents a set of $A$ internal WIMP-nucleus system velocities, 
$A-1$ of which involve the relative
coordinates of bound nucleons (the Jacobi velocities), and one of which is the 
velocity of the DM particle relative to the nuclear center of mass,
\begin{eqnarray}
\vec{v}^\perp &\rightarrow& \left\{ {1 \over 2} \left( \vec{v}_{\chi,\mathrm{in}} + \vec{v}_{\chi,\mathrm{out}}-
\vec{v}_{N,\mathrm{in}}(i)-\vec{v}_{N,\mathrm{out}}(i) \right), i=1,....,A \right\} \nonumber \\
&\equiv&
\vec{v}_T^\perp - \left\{ \vec{\dot{v}}_{N,\mathrm{in}}(i)+ \vec{\dot{v}}_{N,\mathrm{out}}(i), i=1,...,A-1 \right\}.
\end{eqnarray}
The DM particle/nuclear center of mass relative velocity is a c-number,
\be 
\vec{v}_T^\perp ={1 \over 2} \left( \vec{v}_{\chi,\mathrm{in}} + \vec{v}_{\chi,\mathrm{out}}-
\vec{v}_{T,\mathrm{in}}(i)-\vec{v}_{T,\mathrm{out}}(i) \right)
\label{eq:cm}
\ee
while the internal nuclear Jacobi velocities $\vec{\dot{v}}_N$ are operators acting on intrinsic nuclear coordinates.
(That is, for a single-nucleon ($A$=1) target, $\vec{v}^\perp_T \equiv \vec{v}^\perp$, while for all nuclear
targets, there are $A-1$ additional velocity degrees of freedom associated with the Jacobi
internucleon velocities.) This separation is discussed in more detail in the Appendix.

In analogy with Eq. (\ref{eq:simpledensities}) one then obtains the WIMP-nucleus interaction
\begin{eqnarray}
   \sum_{\tau=0,1} \left[
  l_0^\tau~ \sum_{i=1}^A  ~  e^{-i \vec{q} \cdot \vec{x}_i} 
~+~l_0^{A\tau}~ \sum_{i=1}^A ~ {1 \over 2M} \left(-{1 \over i} \overleftarrow{\nabla}_i \cdot  \vec{\sigma}(i) e^{-i \vec{q} \cdot \vec{x}_i}  +  e^{-i \vec{q} \cdot \vec{x}_i}  \vec{\sigma}(i)  \cdot  {1 \over i} \overrightarrow{\nabla}_i \right) \right.  \nonumber \\
 ~+~ \vec{l}_5^\tau \cdot  \sum_{i=1}^A  ~\vec{\sigma}(i)  e^{-i \vec{q} \cdot \vec{x}_i}  
 ~+~ \vec{l}_M^\tau \cdot  \sum_{i=1}^A  ~ {1 \over 2M} \left(-{1 \over i} \overleftarrow{\nabla}_i  e^{-i \vec{q} \cdot \vec{x}_i}  +  e^{-i \vec{q} \cdot \vec{x}_i} {1 \over i} \overrightarrow{\nabla}_i \right)  ~~~~~~~\nonumber \\
~+~ \vec{l}_E^\tau \left. \cdot  \sum_{i=1}^A ~ {1 \over 2M} \left( \overleftarrow{\nabla}_i \times \vec{\sigma}(i)  e^{-i \vec{q} \cdot \vec{x}_i}  +  e^{-i \vec{q} \cdot \vec{x}_i}   \vec{\sigma}(i) \times \overrightarrow{\nabla}_i \right)   \right]_{int} t^\tau(i) ~~~~~~~~~~~~~~~~
\label{eq:fulldensities}
\end{eqnarray}
where the subscript $int$ instructs one to take the intrinsic part of the nuclear operators (that is, the part dependent on the internal Jacobi
velocities).  Comparing to Eq. (\ref{eq:simpledensities}), one sees that three new
velocity-dependent densities appear -- the nuclear axial charge operator, familiar as the
$\beta$ decay operator that mediates $0^+ \leftrightarrow 0^-$ decays; the convection current,
familiar from electromagnetism; and a spin-velocity current that is less commonly discussed,
but does arise as a higher-order correction in weak interactions.  The associated WIMP tensors contain the 
EFT input
\begin{eqnarray}
\label{eq:ls}
l_0^\tau &=& c_1^\tau + i  ( {\vec{q} \over m_N}  \times \vec{v}_T^\perp) \cdot  \vec{S}_\chi  ~c_5^\tau
+ \vec{v}_T^\perp \cdot \vec{S}_\chi  ~c_8^\tau + i {\vec{q} \over m_N} \cdot \vec{S}_\chi ~c_{11}^\tau \nonumber \\
l_0^{A \tau} &=& -{1 \over 2}  \left[ c_7 ^\tau  +i {\vec{q} \over m_N} \cdot \vec{S}_\chi~ c_{14}^\tau \right] \nonumber \\
\vec{l}_5 &=& {1 \over 2} \left[ i {\vec{q} \over m_N} \times \vec{v}_T^\perp~ c_3^\tau + \vec{S}_\chi ~c_4^\tau
+  {\vec{q} \over m_N}~{\vec{q} \over m_N} \cdot \vec{S}_\chi ~c_6^\tau 
+   \vec{v}_T^\perp ~c_7^\tau + i {\vec{q} \over m_N} \times \vec{S}_\chi ~c_9^\tau + i {\vec{q} \over m_N}~c_{10}^\tau \right. \nonumber \\
 && \left.  +  \vec{v}_T^\perp \times \vec{S}_\chi ~c_{12}^\tau  
+i  {\vec{q} \over m_N} \vec{v}_T^\perp \cdot \vec{S}_\chi ~c_{13}^\tau+i \vec{v}_T^\perp {\vec{q} \over m_N} \cdot \vec{S}_\chi ~ c_{14}^\tau+{\vec{q} \over m_N} \times \vec{v}_T^\perp~ {\vec{q} \over m_N} \cdot \vec{S}_\chi ~ c_{15}^\tau  \right]\nonumber \\
\vec{l}_M &=&   i {\vec{q} \over m_N}  \times \vec{S}_\chi ~c_5^\tau - \vec{S}_\chi ~c_8^\tau \nonumber \\
\vec{l}_E &=& {1 \over 2} \left[  {\vec{q} \over m_N} ~ c_3^\tau +i \vec{S}_\chi~c_{12}^\tau - {\vec{q} \over  m_N} \times\vec{S}_\chi  ~c_{13}^\tau-i {\vec{q} \over  m_N} {\vec{q} \over m_N} \cdot \vec{S}_\chi  ~c_{15}^\tau \right]
\end{eqnarray}

In the Appendix the products of plane waves and scalar/vector operators appearing in Eq.  (\ref{eq:fulldensities}) are expanded in spherical and vector spherical harmonics, and the
resulting amplitude is squared, averaged over initial spins and summed over final spins.  
One obtains
\allowdisplaybreaks
\begin{eqnarray}
\label{eq:Ham}
&&{1 \over 2j_\chi + 1} {1 \over 2j_N + 1} \sum_\mathrm{spins} |\CM|^2_\mathrm{nucleus/EFT}  =  {4 \pi \over 2j_N + 1} 
\sum_{ \tau=0,1} \sum_{\tau^\prime = 0,1}  \nonumber \\
&& \left\{  \sum_{J=0,2,...}^\infty    ~\left[ R_{M}^{\tau \tau^\prime}(\vec{v}_T^{\perp 2}, {\vec{q}^{~2} \over m_N^2})   \langle j_N ||~ M_{J;\tau} (q)~ || j_N \rangle 
\langle j_N ||~ M_{J;\tau^\prime} (q)~ || j_N \rangle \right. \right. \nonumber \\
&&~~~+ {\vec{q}^{~2} \over m_N^2} ~ R_{\Phi^{\prime \prime}}^{\tau \tau^\prime}(\vec{v}_T^{\perp 2}, {\vec{q}^{~2} \over m_N^2})     ~\langle j_N ||~\Phi^{\prime \prime}_{J; \tau}(q)~ || j_N \rangle  \langle j_N ||~\Phi^{\prime \prime}_{J; \tau^\prime}(q)~ || j_N \rangle \nonumber \\
&&~~~+  \left.  {\vec{q}^{~2} \over m_N^2} ~  R_{ \Phi^{\prime \prime}M}^{\tau \tau^\prime}(\vec{v}_T^{\perp 2}, {\vec{q}^{~2} \over m_N^2})  ~ \langle j_N ||~ \Phi^{\prime \prime }_{J;\tau}(q)~ || j_N \rangle \langle j_N ||~ M_{J;\tau^\prime} (q) ~|| j_N \rangle \right]  ~~ \nonumber \\
&&  +  \sum_{J=2,4,...}^\infty \left[ {\vec{q}^{~2} \over m_N^2} ~ R_{\tilde{\Phi}^\prime}^{\tau \tau^\prime}(\vec{v}_T^{\perp 2}, {\vec{q}^{~2} \over m_N^2})   ~  \langle j_N || ~\tilde{\Phi}_{J; \tau}^\prime (q)~ || j_N \rangle  \langle j_N ||  ~ \tilde{\Phi}^{\prime}_{J; \tau^\prime}(q)~ || j_N \rangle \right] \nonumber \\
&&  +  \sum_{J=1,3,...}^\infty  \left[ R_{\Sigma^{\prime \prime}}^{\tau \tau^\prime}(\vec{v}_T^{\perp 2}, {\vec{q}^{~2} \over m_N^2})    ~
\langle j_N ||~ \Sigma^{\prime \prime}_{J; \tau}(q)~ || j_N \rangle
 \langle j_N ||~ \Sigma^{\prime \prime}_{J; \tau^\prime}(q)~ || j_N \rangle \right.  \nonumber \\
 &&~~~ +   R_{\Sigma^\prime}^{\tau \tau^\prime}(\vec{v}_T^{\perp 2}, {\vec{q}^{~2} \over m_N^2})  ~  \langle j_N ||~  \Sigma_{J;\tau}^\prime (q)~ || j_N \rangle  \langle j_N || ~\Sigma_{J;\tau^\prime}^{\prime} (q)~ || j_N \rangle \nonumber \\ 
&&~~~ +  {\vec{q}^{~2} \over m_N^2} ~ R_{\Delta}^{\tau \tau^\prime}(\vec{v}_T^{\perp 2}, {\vec{q}^{~2} \over m_N^2})   ~ \langle j_N ||~ \Delta_{J;\tau} (q)~ || j_N \rangle \langle j_N ||~ \Delta_{J; \tau^\prime}(q)~ || j_N \rangle \nonumber \\
&&~~~\left. +{\vec{q}^{~2} \over m_N^2}   R_{\Delta \Sigma^\prime}^{\tau \tau^\prime}(\vec{v}_T^{\perp 2}, {\vec{q}^{~2} \over m_N^2})   ~
\langle j_N || ~ \Delta_{J;\tau} (q) ~|| j_N \rangle  \langle j_N || ~\Sigma^{\prime}_{J; \tau^\prime}(q)~ || j_N \rangle \right]  \Bigg\}  .
\end{eqnarray}
Note that five of the eight terms above are accompanied by a factor of $\vec{q}^{\,2}/m_N^2$.  This
is the parameter identified in Sec. \ref{sec:sec2pt3} that governs the enhancement of the 
composite operators with respect to the point operators for those $\CO_i$ where composite
operators contribute.  Thus one can read off those response functions that are
generated by composite operators from this factor. 
The DM particle response functions are determined by the $c_i^\tau$s,
\allowdisplaybreaks
\begin{eqnarray}
\label{eq:HamC}
 R_{M}^{\tau \tau^\prime}(\vec{v}_T^{\perp 2}, {\vec{q}^{\,2} \over m_N^2}) &=& c_1^\tau c_1^{\tau^\prime } + {j_\chi (j_\chi+1) \over 3} \left[ {\vec{q}^{\,2} \over m_N^2} \vec{v}_T^{\perp 2} c_5^\tau c_5^{\tau^\prime }+\vec{v}_T^{\perp 2}c_8^\tau c_8^{\tau^\prime }
+ {\vec{q}^{\,2} \over m_N^2} c_{11}^\tau c_{11}^{\tau^\prime } \right] \nonumber \\
 R_{\Phi^{\prime \prime}}^{\tau \tau^\prime}(\vec{v}_T^{\perp 2}, {\vec{q}^{\,2} \over m_N^2}) &=& {\vec{q}^{\,2} \over 4 m_N^2} c_3^\tau c_3^{\tau^\prime } + {j_\chi (j_\chi+1) \over 12} \left( c_{12}^\tau-{\vec{q}^{\,2} \over m_N^2} c_{15}^\tau\right) \left( c_{12}^{\tau^\prime }-{\vec{q}^{\,2} \over m_N^2}c_{15}^{\tau^\prime} \right)  \nonumber \\
 R_{\Phi^{\prime \prime} M}^{\tau \tau^\prime}(\vec{v}_T^{\perp 2}, {\vec{q}^{\, 2} \over m_N^2}) &=&  c_3^\tau c_1^{\tau^\prime } + {j_\chi (j_\chi+1) \over 3} \left( c_{12}^\tau -{\vec{q}^{\,2} \over m_N^2} c_{15}^\tau \right) c_{11}^{\tau^\prime } \nonumber \\
  R_{\tilde{\Phi}^\prime}^{\tau \tau^\prime}(\vec{v}_T^{\perp 2}, {\vec{q}^{\,2} \over m_N^2}) &=&{j_\chi (j_\chi+1) \over 12} \left[ c_{12}^\tau c_{12}^{\tau^\prime }+{\vec{q}^{\,2} \over m_N^2}  c_{13}^\tau c_{13}^{\tau^\prime}  \right] \nonumber \\
   R_{\Sigma^{\prime \prime}}^{\tau \tau^\prime}(\vec{v}_T^{\perp 2}, {\vec{q}^{\,2} \over m_N^2})  &=&{\vec{q}^{\,2} \over 4 m_N^2} c_{10}^\tau  c_{10}^{\tau^\prime } +
  {j_\chi (j_\chi+1) \over 12} \left[ c_4^\tau c_4^{\tau^\prime} + \right.  \nonumber \\
 && \left. {\vec{q}^{\,2} \over m_N^2} ( c_4^\tau c_6^{\tau^\prime }+c_6^\tau c_4^{\tau^\prime })+
 {\vec{q}^{\,4} \over m_N^4} c_{6}^\tau c_{6}^{\tau^\prime } +\vec{v}_T^{\perp 2} c_{12}^\tau c_{12}^{\tau^\prime }+{\vec{q}^{\,2} \over m_N^2} \vec{v}_T^{\perp 2} c_{13}^\tau c_{13}^{\tau^\prime } \right] \nonumber \\
    R_{\Sigma^\prime}^{\tau \tau^\prime}(\vec{v}_T^{\perp 2}, {\vec{q}^{\,2} \over m_N^2})  &=&{1 \over 8} \left[ {\vec{q}^{\,2} \over  m_N^2}  \vec{v}_T^{\perp 2} c_{3}^\tau  c_{3}^{\tau^\prime } + \vec{v}_T^{\perp 2}  c_{7}^\tau  c_{7}^{\tau^\prime }  \right]
       + {j_\chi (j_\chi+1) \over 12} \left[ c_4^\tau c_4^{\tau^\prime} +  \right.\nonumber \\
       &&\left. {\vec{q}^{\,2} \over m_N^2} c_9^\tau c_9^{\tau^\prime }+{\vec{v}_T^{\perp 2} \over 2} \left(c_{12}^\tau-{\vec{q}^{\,2} \over m_N^2}c_{15}^\tau \right) \left( c_{12}^{\tau^\prime }-{\vec{q}^{\,2} \over m_N^2}c_{15}^{\tau \prime} \right) +{\vec{q}^{\,2} \over 2 m_N^2} \vec{v}_T^{\perp 2}  c_{14}^\tau c_{14}^{\tau^\prime } \right] \nonumber \\
     R_{\Delta}^{\tau \tau^\prime}(\vec{v}_T^{\perp 2}, {\vec{q}^{\,2} \over m_N^2})&=&  {j_\chi (j_\chi+1) \over 3} \left[ {\vec{q}^{\,2} \over m_N^2} c_{5}^\tau c_{5}^{\tau^\prime }+ c_{8}^\tau c_{8}^{\tau^\prime } \right] \nonumber \\
 R_{\Delta \Sigma^\prime}^{\tau \tau^\prime}(\vec{v}_T^{\perp 2}, {\vec{q}^{\,2} \over m_N^2})&=& {j_\chi (j_\chi+1) \over 3} \left[c_{5}^\tau c_{4}^{\tau^\prime }-c_8^\tau c_9^{\tau^\prime} \right].
\end{eqnarray}

The six nuclear operators appearing in Eq. (\ref{eq:Ham}), familiar from standard-model electroweak interaction theory, are constructed
from the Bessel spherical harmonics
and vector spherical harmonics,
$M_{JM}(q \vec{x}) \equiv j_J(q x) Y_{JM}(\Omega_x)$ and $\vec{M}_{JL}^M \equiv j_L(q x) \vec{Y}_{JLM}(\Omega_x)$,
\allowdisplaybreaks
\begin{eqnarray}
\label{eq:operators}
M_{JM;\tau}(q) &\equiv& \sum_{i=1}^A M_{JM}(q \vec{x}_i)~ t^\tau(i) \nonumber \\
\Delta_{JM;\tau}(q ) &\equiv&\sum_{i=1}^A \vec{M}_{JJ}^M(q \vec{x}_i) \cdot {1 \over q} \vec{\nabla}_i 
~t^\tau(i) \nonumber \\
\Sigma^\prime_{JM;\tau}(q) &\equiv& -i \sum_{i=1}^A \left\{ {1 \over q} \vec{\nabla}_i \times \vec{M}_{JJ}^M (q \vec{x}_i) \right\} \cdot \vec{\sigma}(i)~t^\tau(i) \nonumber \\ 
&=& \sum_{i=1}^A  \left\{ -\sqrt{{J \over 2J+1}}~\vec{M}_{JJ+1}^M(q \vec{x}_i) + \sqrt{{J+1 \over 2J+1}}~\vec{M}_{JJ-1}^M(q \vec{x}_i) \right\} \cdot \vec{\sigma}(i) ~t^\tau(i)\nonumber \\
 \Sigma^{\prime \prime}_{JM;\tau}(q) &\equiv& \sum_{i=1}^A  \left\{ {1 \over q} \vec{\nabla}_i ~ M_{JM} (q \vec{x}_i) \right\} \cdot \vec{\sigma}(i)~t^\tau(i) \nonumber \\
  &=& \sum_{i=1}^A  \left\{ \sqrt{{J+1 \over 2J+1}}~\vec{M}_{JJ+1}^M(q \vec{x}_i) + \sqrt{{J \over 2J+1}}~\vec{M}_{JJ-1}^M(q \vec{x}_i) \right\} \cdot \vec{\sigma}(i) ~t^\tau(i)\nonumber  \\
\tilde{\Phi}^{\prime}_{JM;\tau}(q) &\equiv& \sum_{i=1}^A \left[ \left( {1 \over q} \vec{\nabla}_i \times \vec{M}_{JJ}^M(q \vec{x}_i) \right) \cdot \left(\vec{\sigma}(i) \times {1 \over q} \vec{\nabla}_i \right) + {1 \over 2} \vec{M}_{JJ}^M(q \vec{x}_i) \cdot \vec{\sigma}(i) \right]~t^\tau(i) \nonumber \\
\Phi^{\prime \prime}_{JM;\tau}(q ) &\equiv& i  \sum_{i=1}^A\left( {1 \over q} \vec{\nabla}_i  M_{JM}(q \vec{x}_i) \right) \cdot \left(\vec{\sigma}(i) \times {1 \over q} \vec{\nabla}_i \right)~t^\tau(i)
\end{eqnarray}
Equations (\ref{eq:Ham}), (\ref{eq:HamC}), and (\ref{eq:operators}) comprise the general
expression for the WIMP-nucleon spin-averaged transition probability.  $M,~\Delta,~\Sigma^\prime,
~\Sigma^{\prime \prime},~ \tilde{\Phi}^\prime,~\mathrm{and}~\Phi^{\prime \prime}$ transform as
vector charge, vector transverse magnetic, axial transverse electric, axial longitudinal, vector
transverse electric, and vector longitudinal operators, respectively.  These are the allowed responses under the assumption that the nuclear ground state is an approximate eigenstate of P and CP,
and thus we have derived the most general form of the cross section.

As we will discuss in more detail in Sec. \ref{sec:sec5}, our Mathematica script assumes that the
nuclear wave functions are of the standard shell model form -- expanded over a set Slater determinants --
where the underlying single-particle basis is the harmonic oscillator.  
In that case Eq. (\ref{eq:Ham}) gives the cross section as a sum of products of WIMP 
$ R_{k}^{\tau \tau^\prime}(\vec{v}_T^{\perp 2}, {\vec{q}^{\,2} \over m_N^2})$ and
nuclear $W_k^{\tau \tau^\prime}(y)$ response functions, where $y=(qb/2)^2$ with $b$
the harmonic oscillator size parameter.  That is, the evolution of the nuclear
responses with $q$ is determined by the single dimensionless parameter $y$. 
Eq. (\ref{eq:Ham}) can then be written compactly as
\allowdisplaybreaks
\begin{eqnarray}
\label{eq:HamHO}
&&{1 \over 2j_\chi + 1} {1 \over 2j_N + 1} \sum_\mathrm{spins} |\CM|^2_\mathrm{nucleus-HO/EFT}  =  {4 \pi \over 2j_N + 1} 
\sum_{ \tau=0,1} \sum_{\tau^\prime = 0,1}  \nonumber \\
&& \left\{  ~\left[ R_{M}^{\tau \tau^\prime}(\vec{v}_T^{\perp 2}, {\vec{q}^{~2} \over m_N^2})~W_{M}^{\tau \tau^\prime}(y) + R_{\Sigma^{\prime \prime}}^{\tau \tau^\prime}(\vec{v}_T^{\perp 2}, {\vec{q}^{~2} \over m_N^2})   ~W_{\Sigma^{\prime \prime}}^{\tau \tau^\prime}(y)  +   R_{\Sigma^\prime}^{\tau \tau^\prime}(\vec{v}_T^{\perp 2}, {\vec{q}^{~2} \over m_N^2}) ~ W_{\Sigma^\prime}^{\tau \tau^\prime}(y) \right] \right. \nonumber \\  
&&+ {\vec{q}^{~2} \over m_N^2} ~\left[ R_{\Phi^{\prime \prime}}^{\tau \tau^\prime}(\vec{v}_T^{\perp 2}, {\vec{q}^{~2} \over m_N^2}) ~ W_{\Phi^{\prime \prime}}^{\tau \tau^\prime}(y)  
+  R_{ \Phi^{\prime \prime}M}^{\tau \tau^\prime}(\vec{v}_T^{\perp 2}, {\vec{q}^{~2} \over m_N^2})  ~W_{ \Phi^{\prime \prime}M}^{\tau \tau^\prime}(y) +   R_{\tilde{\Phi}^\prime}^{\tau \tau^\prime}(\vec{v}_T^{\perp 2}, {\vec{q}^{~2} \over m_N^2})   W_{\tilde{\Phi}^\prime}^{\tau \tau^\prime}(y) \right. \nonumber \\
&&~~~+ \left. \left.  R_{\Delta}^{\tau \tau^\prime}(\vec{v}_T^{\perp 2}, {\vec{q}^{~2} \over m_N^2}) ~ W_{\Delta}^{\tau \tau^\prime}(y) 
 +  R_{\Delta \Sigma^\prime}^{\tau \tau^\prime}(\vec{v}_T^{\perp 2}, {\vec{q}^{~2} \over m_N^2})  ~W_{\Delta \Sigma^\prime}^{\tau \tau^\prime}(y)   \right]  \right\}  
\end{eqnarray}
where
\begin{eqnarray}
\label{eq:nucresponse}
 W_{M}^{\tau \tau^\prime}(y)&=& \sum_{J=0,2,...}^\infty    \langle j_N ||~ M_{J;\tau} (q)~ || j_N \rangle 
\langle j_N ||~ M_{J;\tau^\prime} (q)~ || j_N \rangle \nonumber \\
 W_{\Sigma^{\prime \prime}}^{\tau \tau^\prime}(y) &=&
\sum_{J=1,3,...}^\infty \langle j_N ||~ \Sigma^{\prime \prime}_{J; \tau}(q)~ || j_N \rangle
 \langle j_N ||~ \Sigma^{\prime \prime}_{J; \tau^\prime}(q)~ || j_N \rangle \nonumber \\
 W_{\Sigma^\prime}^{\tau \tau^\prime}(y)  &=&\sum_{J=1,3,...}^\infty  \langle j_N ||~  \Sigma_{J;\tau}^\prime (q)~ || j_N \rangle  \langle j_N || ~\Sigma_{J;\tau^\prime}^{\prime} (q)~ || j_N \rangle \nonumber \\ 
W_{ \Phi^{\prime \prime}}^{\tau \tau^\prime}(y) &=& \sum_{J=0,2,...}^\infty     ~\langle j_N ||~\Phi^{\prime \prime}_{J; \tau}(q)~ || j_N \rangle  \langle j_N ||~\Phi^{\prime \prime}_{J; \tau^\prime}(q)~ || j_N \rangle \nonumber \\
 W_{ \Phi^{\prime \prime}M}^{\tau \tau^\prime}(y)  &=&  \sum_{J=0,2,...}^\infty  \langle j_N ||~ \Phi^{\prime \prime }_{J;\tau}(q)~ || j_N \rangle \langle j_N ||~ M_{J;\tau^\prime} (q) ~|| j_N \rangle  \nonumber \\
 W_{\tilde{\Phi}^\prime}^{\tau \tau^\prime}(y)  &=&  \sum_{J=2,4,...}^\infty ~  \langle j_N || ~\tilde{\Phi}_{J; \tau}^\prime (q)~ || j_N \rangle  \langle j_N ||  ~ \tilde{\Phi}^{\prime}_{J; \tau^\prime}(q)~ || j_N \rangle \nonumber \\
 W_{\Delta}^{\tau \tau^\prime}(y)  &=&\sum_{J=1,3,...}^\infty  \langle j_N ||~ \Delta_{J;\tau} (q)~ || j_N \rangle \langle j_N ||~ \Delta_{J; \tau^\prime}(q)~ || j_N \rangle \nonumber \\
W_{\Delta \Sigma^\prime}^{\tau \tau^\prime}(y)  &=&\sum_{J=1,3,...}^\infty 
\langle j_N || ~ \Delta_{J;\tau} (q) ~|| j_N \rangle  \langle j_N || ~\Sigma^{\prime}_{J; \tau^\prime}(q)~ || j_N \rangle.
\end{eqnarray}
Equations (\ref{eq:HamHO}), (\ref{eq:HamC}), and (\ref{eq:nucresponse}) are the key formulas
evaluated by the Mathematica script of Sec. \ref{sec:sec7}.  Parity and CP restrict the sums over multipolarities $J$ to only even or only odd terms, depending on the transformation properties of the operators,
again as described in the Appendix.

The physics of these six nuclear response functions is more easily seen by examining the long-wavelength
forms of the corresponding operators.  The operators that are nonvanishing as $q \rightarrow 0$ are
\allowdisplaybreaks
\begin{eqnarray}
\label{eq:operators_limit}
\sqrt{4 \pi} M_{00;\tau}(0) &=& \sum_{i=1}^A~ t^\tau(i)  \nonumber \\
\sqrt{4 \pi} \Delta_{1M;\tau}(0) &=&-{1 \over \sqrt{6}}  \sum_{i=1}^A~ l_{1M}(i)~ t^\tau(i) 
 \nonumber \\
\sqrt{4 \pi} \Sigma^\prime_{1M;\tau}(0 ) &=& \sqrt{{2 \over 3}} \sum_{i=1}^A~ \sigma_{1M}(i)~ t^\tau(i) \nonumber \\
\sqrt{4 \pi}  \Sigma^{\prime \prime}_{1M;\tau}(0) &=&{1 \over  \sqrt{3}} \sum_{i=1}^A~ \sigma_{1M}(i)~ t^\tau(i) \nonumber  \\
\sqrt{4 \pi}\tilde{\Phi}^{\prime}_{2M;\tau}(0) &=& -{1 \over \sqrt{5}} \sum_{i=1}^A~\left[ x(i) \otimes \left(\vec{\sigma}(i) \times {1 \over i} \vec{\nabla}(i) \right)_1 \right]_2 t^\tau(i) \nonumber \\
\sqrt{4 \pi} \Phi^{\prime \prime}_{JM; \tau}(0 ) &=&  \left\{ \begin{array}{lr} {1 \over 3} \sum\limits_{i=1}^A ~ \vec{\sigma}(i) \cdot \vec{l}(i)~ t^\tau(i) & J=0 \\ ~ & ~ \\
 -{1 \over \sqrt{5}} \sum\limits_{i=1}^A ~\left[ x(i) \otimes \left(\vec{\sigma}(i) \times {1 \over i} \vec{\nabla}(i) \right)_1 \right]_2 t^\tau(i) & ~~~~ J=2 \end{array} \right.
\end{eqnarray}
where the operator $\Phi^{\prime \prime}$ has scalar and tensor components that survive.
Two combinations of operators are, of course, related
to the spin-independent/spin-dependent forms
\begin{eqnarray}
|M_{F;\tau}^N(0)|^2  &\equiv& {4 \pi \over 2j_N+1} | \langle j_N ||  M_{0;\tau}(0)  ||j_N\rangle|^2 \nonumber \\
|M_{GT;\tau}^N(0)|^2 &\equiv& {4 \pi \over 2j_N+1} \left(  |\langle j_N || \Sigma^{\prime \prime}_{1;\tau}(0) || j_N \rangle |^2 + |\langle j_N || \Sigma^\prime_{1;\tau}(q) || j_N \rangle|^2 \right)
\end{eqnarray}

In the next section we will describe in more detail some of the differences between
this form and the point-nucleus and allow forms, where the only the simple
Fermi and Gamow-Teller operators arise.  But one can make some initial observations here:
\begin{itemize}
\item The most general form of the WIMP-nucleus elastic scattering probability has six,
not two, response functions.  They are associated the squares of the matrix elements of the six operators given in
Eqs. (\ref{eq:operators}).  There are also two interference terms
($\Phi^{\prime \prime} \leftrightarrow M \mathrm{~and~} \Delta \leftrightarrow \Sigma^\prime$).  
\item The spin response familiar from the standard allowed treatment of WIMP-nucleus 
interactions splits into separate longitudinal and transverse components, as various
candidate effective interactions do not couple to all spin projections symmetrically.
The associated operators, $\Sigma^{\prime \prime}$ and $\Sigma^{\prime}$, are
proportional in the long-wavelength limit, but are distinct at finite $\vec{q}^{\,2}$ because their
associated form factors differ.
\item Three new response functions are generated from couplings to the intrinsic
velocities of nucleons, and consequently reflect the composite nature of the nucleus.
Reflecting their finite-nuclear-size origin,  the three responses appear in Eq. (\ref{eq:Ham}) with an explicit factor of $\vec{q}^{\,2}/m_N^2$.
\item Two scalar responses appear in Eq. (\ref{eq:Ham}), generated by the standard
Fermi operator $1(i)$ and by the new  spin-orbit operator $\vec{\sigma} \cdot \vec{l}(i)$.  Thus
both are ``spin-independent" responses - responses associated with operators that transform
as scalars under rotations.
\item There are three vector responses, two associated with the (in general, independent) longitudinal and transverse
projections of spin and the third with the orbital angular momentum operator $\vec{l}(i)$.
These three operators transform under rotations as $\vec{j}_N$, and all 
thus require a nuclear ground
state spin of $j_N \ge 1/2$.  It was shown in \cite{nonrelEFT} that among the various nuclear
targets now in use for dark matter studies, the relative strength of spin and orbital 
transition probabilities can differ by two orders magnitude or more.
\item One response function, generated by $\tilde{\Phi}^\prime$, is
tensor, and thus only contributes if $j_N \ge 1$.  This response function is somewhat exotic,
coming from interactions  $\CO_{12}$, $\CO_{13}$, and $\CO_{15}$ that we have noted do not
arise for traditional spin-0 or spin-1 exchanges.   
\item The EFT result of Eq. (\ref{eq:Ham}) and the spin-dependent/spin-independent result
of Eq. (\ref{eq:allowed}) coincide if one takes $\vec{q}^{\,2} \rightarrow 0$ and also $\vec{v}_T^{\perp 2} \rightarrow 0$, a limit that zeros out all contributions from low-energy constants other that $c_1$ and $c_4$.
But away from this limit they differ.  This illustrates the inconsistency of the standard spin-independent/
spin-dependent formulation with form factors: one selectively includes
powers of $\vec{q} \cdot \vec{x}(i)$ to modified the Fermi $1(i)$ and $\vec{\sigma}(i)$
operators through form factors, while not using those same factors to create new operators. 
\end{itemize}

\section{Experiment: Cross Sections and Rates}
 \label{sec:sec4}
In this section we present the basic formulas for cross sections and event rates --
the quantities of interest to experimentalists -- in terms
of the square of the (Galilean) invariant amplitude.  We discuss EFT power counting and 
its implications in the context of the dependence of the total cross section $\sigma(v)$ of
the relevant physical parameters.
\subsection{Differential cross sections and rates}
The cross section for
WIMP scattering off a nucleus in the laboratory frame is obtained by folding the 
transition probability with the corresponding Lorentz-invariant phase space,
\begin{equation}
d \sigma = {1 \over v} {m_\chi \over  E_\chi^i} 
\left[ {1 \over 2j_\chi + 1} {1 \over 2j_N + 1} \sum_\mathrm{spins} |\CM|^2 \right] 
{m_\chi \over E_\chi^f} ~ {d^3p^\prime \over (2 \pi)^3} {m_T \over E_T^f} {d^3k^\prime \over (2 \pi)^3}  (2 \pi)^4 \delta^4(p+k-p^\prime-k^\prime)
\label{eq:M}
\end{equation}
where $p,p^\prime$ and $k,k^\prime$ are the initial and final dark-matter particle and 
nuclear momenta.  $\CM$, in most other applications the Lorentz invariant amplitude, is
in our construction the Galilean invariant amplitude, due to the nonrelativistic nature of the 
scattering.  As this expression is in the lab frame, $v$ is the initial WIMP velocity; the target
is at rest.

$\CM$ is a function of $v$ and one other variable.  If we define a scattering angle by the
direction of nuclear recoil relative to the initial WIMP velocity, $\hat{v} \cdot \hat{k}^\prime
=- \hat{v} \cdot \hat{q}= \cos{\theta}$,
then that second variable can be taken to be 
$\vec{q}^{\,2}$, or equivalently the energy of the recoiling nucleus
$E_R = \vec{q}^{\,2}/2 m_T$, or equivalently, using the lab-frame energy conservation
condition 
\be 
{\vec{p}^{\,2} \over 2 m_\chi} -{ (\vec{p}-\vec{k}^\prime)^2 \over 2 m_\chi} - {\vec{k}^{\prime\,2} \over 2 m_T}=0 ~\Rightarrow~ {\vec{k}^{\prime \,2} \over 2 \mu_T} = \vec{v} \cdot \vec{k}^\prime ~\rightarrow~ {\vec{k}^{\prime \,2} \over 4 \mu_T^2 v^2} = \cos^2{\theta}~\Rightarrow {\vec{q}^2 \over 2 \mu_T^2 v^2} = 1+ \cos{2 \theta},
\ee
the angular variable $\cos{2 \theta}$.   Note that as $\vec{v} \cdot \vec{k}^\prime \ge 0$, $0 \le \theta \le \pi/2$, and thus
$0 \le 2 \theta \le  \pi$.  We can integrate Eq. (\ref{eq:M}) to obtain the differential cross sections
\be
{d \sigma(v,E_R) \over dE_R} = 2 m_T {d \sigma(v,\vec{q}^{\,2}) \over d\vec{q}^{\,2}} =2m_T  {1 \over 4 \pi v^2}  \left[ {1 \over 2j_\chi + 1} {1 \over 2j_N + 1} \sum_\mathrm{spins} |\CM^{Nuc}|^2 \right] 
\ee
\be
{d \sigma(v,\theta) \over d \cos{2 \theta}} = 2 \mu_T^2 v^2 {d \sigma(v,\vec{q}^{\,2}) \over d\vec{q}^{\,2}} = {\mu_T^2 \over 2 \pi} \left[ {1 \over 2j_\chi + 1} {1 \over 2j_N + 1} \sum_\mathrm{spins} |\CM^{Nuc}|^2 \right] 
\ee

The differential scattering rate per detector and per target nucleus averaging over the
galactic WIMP velocity distribution can then be calculated
\begin{eqnarray}
 {d R_D \over dE_R} = N_T {d R_T \over dE_R} = N_T  \int {d \sigma(v,E_R) \over dE_R}  v dn_\chi &=& N_T n_\chi  \int _{v>v_\mathrm{min}} {d \sigma(v,E_R) \over dE_R}  v f_E(\vec{v}) d^3v \nonumber \\
 &\equiv& N_T n_\chi \Bigl\langle v  {d \sigma(v,E_R) \over dE_R}\Bigr\rangle_{v>v_\mathrm{min}}
 \end{eqnarray}
where $N_T$ is the number of target nuclei in the detector, $n_\chi$ is the local
number density of dark matter particles, and $f_E(\vec{v})$ the normalized velocity
distribution of the dark matter particles.  Thus $n_\chi= \rho_\chi/m_\chi$ where $\rho_\chi$ is
the dark matter density.  The integral over velocities begins with the minimum velocity
required to produce a recoil energy $E_R$,
\be 
v_\mathrm{min}= v_\mathrm{min}(E_R) ={q \over 2 \mu_T} = {1 \over \mu_T} \sqrt{{m_T E_R \over 2}} .
\ee
Similarly,
\begin{eqnarray}
 {d R_D \over d \cos{2 \theta}} =N_T {d R_T \over d \cos{2 \theta}}  =N_T n_\chi  \int  {d \sigma(v,E_R) \over d \cos{2 \theta}}  v f_E(\vec{v}) d^3v
 &\equiv& N_T n_\chi \Bigl\langle v  {d \sigma(v,E_R) \over d \cos{2 \theta}}\Bigr\rangle.
 \end{eqnarray}
Here there is no restriction on the recoil energy, and thus no requirement for a minimum velocity.

In the same way,  one can calculate the total cross section
\be 
\sigma(v) = \int_0^{4 v^2 \mu_T^2}  {d \sigma(v,\vec{q}^{\,2}) \over d \vec{q}^{\,2}}~d\vec{q}^{\,2}.
\ee
The total scattering rate per detector $R_D$ and per target nucleus $R_T$ become
\be
R_D = N_T R_T = N_T n_\chi \int  \sigma(v) v f_E(\vec{v}) d^3v \equiv N_T n_\chi \bigl\langle v   \sigma(v) \bigr\rangle.
\ee

\subsection{Experimental output of the Mathematica script}
The following quantities can be obtained directly from the script, depending on the options
chosen.  The evaluations can be done for a specific target nuclei --
e.g., a nucleus with a definite (N,Z) -- or for a natural target, by summing over $N$ for fixed $Z$,
weighting each component by the natural abundance, to obtain an effective cross section per
target nucleus.
\begin{enumerate}
\item The laboratory differential cross section
\be
{d\sigma(v,E_R) \over dE_R}
\ee
 for fixed $v$, as functions of $E_R<2 v^2 \mu_T^2/M_T$.
 \item The flux-weighted differential event rate averaged over the normalized galactic WIMP velocity distribution
 \be
\frac{dR_D}{d E_R} &=& N_T \left\< n_\chi v \frac{d \sigma}{d E_R} \right\>
\ee
as functions of $E_R$, $\vec{q}^{\,2}$.
\item The total cross section $\sigma(v)$ as a function of $v$.
\end{enumerate}

\subsection{EFT Power Counting: The Parametric Dependence of Total Cross Sections}
\label{sec:sec4pt3}
An inspection of Eq. (\ref{eq:Ham}) shows that if all operators are evaluated in the long-wavelength
limit (that is, ignoring form factors), the equation reduces to the point-nucleus 
result given in Eq. (\ref{eq:point}), if in addition operators other than 
$M$, $\Sigma^{\prime \prime}$, and $\Sigma^\prime$ are eliminated.   Thus by 
working in the long-wavelength limit, keeping all operators in leading order, one has a
simple test of the relevance of the new operators, those other than the Fermi and
Gamow-Teller ones.  A suitable observable for this comparison is $\sigma(v)$, as the
integration over $\vec{q}^{\,2}$ in Eq. (\ref{eq:Ham}) is easily done using the 
laboratory-frame relation $\vec{v}_T^{\perp 2} = \vec{v}^{\,2} + \vec{q}^{\,2}/4 \mu_T^2$.  One finds
for each of the EFT interactions (and, for simplicity, considering couplings only to protons, 
so that the results match Eq. (\ref{eq:point}))\\
\begin{eqnarray}
\sigma_{c_1^p}(v) &=&     c_1^{p 2}~{\mu_T^2 \over \pi}   \left[ {4 \pi \over 2J_i+1} \langle M_{0;p}(0) \rangle^2 \right] \nonumber \\
\sigma_{c_3^p}(v) &=&    c_3^{p 2}~ v^4~ {\mu_T^2 \over \pi} \left[ {4 \pi \over 2J_i+1}  \left( {\mu_T \over m_N} \right)^2  {1 \over 12} \left( \langle \Sigma^\prime_{1;p}(0) \rangle^2 +16 \left( {\mu_T \over m_N} \right)^2  \left(\langle \Phi_{0;p}^{\prime \prime}(0)\rangle^2+\langle \Phi_{2;p}^{\prime\prime}(0)\rangle^2 \right) \right)\right]\nonumber \\
\sigma_{c_4^p}(v) &=&    c_4^{p 2}~ {\mu_T^2 \over \pi} \left[{4 \pi \over 2J_i +1} S_\chi(S_\chi+1) {1 \over 12} \left(\langle \Sigma^\prime_{1;p} \rangle^2+ \langle \Sigma^{\prime \prime}_{1;p} \rangle^2 \right)\right] \nonumber \\
\sigma_{c_5^p}(v) &=&    c_5^{p 2}~ v^4~{\mu_T^2 \over \pi} \left[{ 4 \pi \over 2J_i+1}  \left( {\mu_T \over m_N} \right)^2 S_\chi(S_\chi+1)  {2 \over 9} \left( \langle M_{0;p} \rangle^2 +8 \left( {\mu_T \over m_N} \right)^2  \langle\Delta_{1;p} \rangle^2 \right) \right]\nonumber \\
\sigma_{c_6^p}(v) &=&  c_6^{p 2}~ v^4~{\mu_T^2  \over \pi} \left[ {4 \pi \over 2J_i+1} \left( {\mu_T \over m_N} \right)^4 S_\chi(S_\chi+1) {4 \over 9} \langle \Sigma^{\prime \prime}_{1;p} \rangle^2 \right] \nonumber \\
\sigma_{c_7^p}(v) &=&   c_7^{p 2}~v^2~{\mu_T^2 \over \pi} \left[ {4 \pi \over 2J_i+1}  {1 \over 16} \langle \Sigma^\prime_{1;p} \rangle^2 \right] \nonumber \\
\sigma_{c_8^p}(v) &=&   c_8^{p 2}~ v^2~ {\mu_T^2\over \pi}  \left[ {4 \pi \over 2J_i+1}  S_\chi(S_\chi+1)  {1 \over 6} \left( \langle M_{0;p} \rangle^2 + 4 \left( {\mu_T \over m_N} \right)^2  \langle\Delta_{1;p} \rangle^2 \right) \right]\nonumber \\
\sigma_{c_9^p}(v) &=&    c_9^{p 2}~v^2 ~{\mu_T^2 \over \pi} \left[{4 \pi \over 2J_i+1}   \left( { \mu_T \over m_N} \right)^2 S_\chi (S_\chi+1) {1 \over 6} \langle \Sigma^\prime_{1;p} \rangle^2 \right] \nonumber \\
\sigma_{c_{10}^p}(v) &=&    c_{10}^{p 2} ~v^2~{\mu_T^2 \over \pi} \left[ {4 \pi \over 2J_i+1}  \left( { \mu_T \over m_N} \right)^2  {1 \over 2} \langle \Sigma^{\prime \prime}_{1;p} \rangle^2 \right] \nonumber \\
\sigma_{c_{11}^p}(v) &=&      c_{11}^{p 2}~v^2~{\mu_T^2  \over  \pi } \left[ {4 \pi  \over 2J_i+1} \left( { \mu_T \over m_N} \right)^2 S_\chi (S_\chi+1) {2 \over 3} \langle M_{0;p} \rangle^2 \right] \nonumber \\
\sigma_{c_{12}^p}(v) &=&  c_{12}^{p 2}~  v^2~{\mu_T^2 \over  \pi } \left[{4 \pi \over 2J_i+1}  S_\chi(S_\chi+1)  {1 \over 24} \Bigg( \langle \Sigma^{\prime \prime}_{1;p} \rangle^2 + {1 \over 2} \langle  \Sigma^\prime_{1;p} \rangle^2 \right. \nonumber \\
&&~~~~~~~~~~~~~~~~~~~~+ \left.  4 \left( {\mu_T \over m_N} \right)^2  \left(\langle \tilde{\Phi}_{2;p}^\prime \rangle^2+\langle \Phi_{0;p}^{\prime \prime}\rangle^2+\langle \Phi_{2;p}^{\prime\prime}\rangle^2 \right) \Bigg) \right] \nonumber \\
\sigma_{c_{13}^p}(v) &=&   c_{13}^{p 2} ~ v^4~{\mu_T^2\over \pi}  \left[ {4 \pi \over 2J_i+1}  \left( {\mu_T \over m_N} \right)^2 S_\chi(S_\chi+1)  {1 \over 18} \left( \langle \Sigma^{\prime \prime}_{1;p} \rangle^2 +8 \left( {\mu_T \over m_N} \right)^2  \langle \tilde{\Phi}_{2;p}^{ \prime}\rangle^2  \right)\right]\nonumber \\
\sigma_{c_{14}^p}(v) &=&  c_{14}^{p 2}~ v^4~{\mu_T^2 \over \pi } \left[ {4 \pi \over 2 J_i+1}  \left( { \mu_T \over m_N} \right)^2 S_\chi(S_\chi+1) {1 \over 36} \langle \Sigma^\prime_{1;p} \rangle^2 \right]  \nonumber \\
\sigma_{c_{15}^p}(v) &=&   c_{15}^{p 2} ~ v^6~{\mu_T^2\over \pi}  \left[ {4 \pi \over 2J_i+1}  \left( {\mu_T \over m_N} \right)^4 S_\chi(S_\chi+1)  {1 \over 18} \left( \langle \Sigma^{\prime}_{1;p} \rangle^2 +24 \left( {\mu_T \over m_N} \right)^2 \left(  \langle {\Phi}_{0;p}^{ \prime \prime}\rangle^2  + {\Phi}_{2;p}^{ \prime \prime}\rangle^2  \right) \right) \right] \nonumber \\
~
\label{eq:simple}
\end{eqnarray}
where we have used $\langle\hat{O}_{J;p} \rangle$ as shorthand for the matrix
element $\langle j_N || \hat{O}_{J;p} || j_N \rangle$.

The pattern one sees in the above results reflects an underlying EFT power counting.
Suppose we designate our WIMP-nucleon operators as $\CO_i(\alpha_i,\beta_i)$ where $\alpha_i$ 
and $\beta_i$ denote
the number of powers of $\vec{v}^\perp$ and $\vec{q}/m_N$, respectively, appearing in the operator,
\be
\CO_i(\alpha_i,\beta_i) \leftrightarrow \left[\vec{v}^\perp\right]^{\alpha_i }~ \left[ {\vec{q} \over m_N} \right]^{\beta_i}~~~~\alpha_i =0,1.
\ee
The total cross section has the form
\be
\sigma_i(v) \sim c_i^2~ \mu_T^2  ~(v^2)^{\alpha_i+\beta_i} \left({\mu_T^2 \over m_N^2}\right)^{\beta_i} \left[a_T^i\langle\hat{O}_i^T \rangle^2
+ a^i_N\delta_{\alpha_i 1}\langle\hat{O}_i^N \rangle^2 \left({\mu_T^2 \over m_N^2}\right)^{\alpha_i}~ \right]
\label{eq:scaling}
\ee
where $\hat{O}_i^T$ and $\hat{O}_I^N$ represent one of the dimensionless point $(M_0, \Sigma^\prime_1,\Sigma^{\prime \prime}_1)$ or composite $(\Delta_1,\tilde{\Phi}_2,\Phi^{\prime \prime}_{0,2})$ operators, respectively, and $a_T^i$
and $a_N^i$ represent simple numerical factors, e.g.,
\be 
a_T^{15} = {S_\chi (S_\chi+1) \over 18}~~~~~~~~a_N^{15} = {2 S_\chi(S_\chi+1) \over 3}~~~~~\mathrm{typically~with}~~{a_N^i \over a_T^i} \sim 10.
\ee

We see that total cross sections and thus total rates depend on the dimensionless parameters $v$ and $\mu_T/m_N$, but that the parametric dependence on $\mu_T/m_N$  depends on the operator
type, point or composite.
The cross section for the composite operators have the simple behavior
\be
\sigma_i(v) \Big|_{N} \sim \left[ v^2 {\mu_T^2 \over m_N^2} \right]^{\alpha_i+\beta_i}.
\ee
where the value of $\alpha_i+\beta_i$=0,1,2,3  is equivalent to our EFT designation LO, NLO, NNLO,
N$^3$LO. This reflects the fact that there are $\alpha_i+\beta_i$ powers of $\vec{q}/m_N$ in the composite
operator, with one factor ($\alpha_i=1$) coming from $i \vec{q} \cdot \vec{x}(i)$ in combination with $\vec{v}_N(i)$.
The cross section contributions of the point-nucleus operator scale as
\be
\sigma_i(v) \Big|_{T} \sim (v^2)^{\alpha_i} ~ \left[ v^2 {\mu_T^2 \over m_N^2} \right]^{\beta_i}.
\ee
There are $\beta_i$ powers of $\vec{q}/m_N$, while the accompany 
velocity is not a nuclear operator, but the c-number $v^\perp_T$.

Both terms are generally present (see the exception below) if there is a velocity coupling.
Consequently the neglect of composite operators for interactions
with derivative couplings not only leads to a cross section that is much
too small (by a factor $\sim (a_N^i/a_T^i)(\mu_T^2/m_N^2)$), but 
produces a cross section with the wrong parametric dependence on $m_T$ and $m_\chi$,
potentially distorting comparisons among experiments that are using different nuclear
targets, as well as sensitivity plots as a function of $m_\chi$.
     
If this calculation is extended to the full operators rather than just there long-wavelength forms,
the two terms comprising Eq. (\ref{eq:scaling}) are modified by factors $F^2_T(\gamma)$ and $F_N^2(\gamma)$, where $\gamma = (b \mu_T v)^2$.  
Thus three dimensionless parameters, $v$, $\gamma$, and $\mu_T/m_N$, describe the total
cross section's dependence on the WIMP velocity,
the nuclear size, and the WIMP-to-nucleus mass ratio, respectively.

\subsection{Comparison of the Standard and EFT Results}
The above results should be helpful to those wanting to understand the limitations of standard
treatments that retain only the Fermi and Gamow-Teller responses.  The consequences
are operator specific:
\begin{enumerate}
\item Operators $\CO_1$ and $\CO_4$ are the simple-minded spin-independent and 
spin-dependent operators.  Their coupling is to total spin and total charge (in the general case,
some combination of N and Z, depending on chosen operator isospin).  These operators are
point operators, and thus the standard treatment is valid in all respects.
\item The coupling of operator $\CO_{11}$ to the nucleus is $1_i$, the vector charge operator.  As the
nuclear physics is identical to that of $\CO_1$, a standard spin-independent analysis 
would correctly model the nuclear physics of this operator.  However, the dependence
of rates on the WIMP velocity distribution differ for $\CO_1$ and $\CO_{11}$, and this
difference would normally not be addressed in comparisons among experiments if only
interaction $\CO_1$ is retained  (see point 7
below).
\item The operators $\CO_6$ and $\CO_{10}$ couple to the nucleus through longitudinal
spin, $\vec{q} \cdot \vec{\sigma}(i)$, while $\CO_9$ couples through transverse spin,
$\vec{q} \times \vec{\sigma}(i)$.   For these operators, the standard analysis based on a
spin-dependent coupling would yield the right threshold ($\vec{q} \rightarrow 0$) coupling
to the nucleus, but misrepresent the form factors (as $\Sigma^{\prime}$ 
and $\Sigma^{\prime \prime}$ are described by distinct form factors).  The predicted dependence of
rates on the galactic WIMP velocity distribution also differs from the standard
$\CO_4$ interaction (see point 7 below).
\item  The operators $\CO_3$, $\CO_5$, $\CO_8$, $\CO_{12}$, $\CO_{13}$, and $\CO_{15}$ involve
velocity-dependent couplings to the nucleus.  The standard spin-independent/spin-dependent
analysis grossly misrepresents the physics of these operators, leading to errors that can
exceed several orders of magnitude.  They couple dominantly through the new composite
operators $\Delta$, $\tilde{\Phi}^\prime$, and $\Phi^{\prime \prime}$:  the contributions of these operators to the 
cross section are parametrically enhanced relative to those of the standard operators 
by the factor  $(4-24) \times (\mu_T/m_N)^2 \sim
10 A^2$.  
 The resulting large errors can be partially
mitigated in the case of $\CO_5$ and $\CO_8$ because the new operators compete
with $M_0$, which can be coherent if isospin couplings are dialed to
make the operator primarily isoscalar.  But even in this favorable case, the error can
be an order of magnitude.
\item  In all of the cases above, the standard treatment would distort the multipolarity of
the coupling.  Operators $\CO_3$, $\CO_{12}$, $\CO_{13}$, and $\CO_{15}$ would appear in the standard
treatment as spin-dependent interactions, coupling through $\Sigma_1^\prime$ and
$\Sigma_1^{\prime \prime}$, and thus could be probed only if the
target has $j_N \ge 1/2$.  In fact, $\CO_3$, $\CO_{12}$, and $\CO_{15}$ have dominant scalar
couplings through $\Phi_0^{\prime \prime}$, which we have noted is proportional
to $\vec{\sigma}(i) \cdot \vec{l}(i)$ -- an operator that is not only scalar, but is 
quasi-coherent, as discussed in \cite{nonrelEFT}.  The dominant 
contribution from $\CO_{13}$ is through the tensor operator $\tilde{\Phi}^{ \prime}_2$,
which requires $j_N \ge 1$, a possibility totally
outside the standard description.  
\item Two operators remain that at first appear puzzling: $\CO_7$ and $\CO_{14}$
have velocity-dependent couplings to the nucleus, but unlike the operators discussing in point 5,
they have standard spin-dependent couplings, and no contribution from the new composite
operators.  This result is a consequence of the good P and CP of nuclear wave functions.
These operators couple to the nucleus through the axial-charge, $\vec{S}_N \cdot \vec{v}^\perp$.
 When one combines
$\vec{S}_N \cdot \vec{v}^\perp$ with $e^{-i \vec{q} \cdot \vec{x}_i}$ to produce
multipole operators in the standard way, the matrix elements of the even multipoles
vanish by parity, while those of the odd multipoles vanish
by CP (or, equivalently, time-reversal invariance).  Consequently all contributions of 
intrinsic velocities to $\CO_7$ and $\CO_{14}$ vanish.  Thus the only contribution to 
the axial charge operator that survives is the single degree-of-freedom corresponding
to the nuclear center-of-mass velocity.  As this velocity is a c-number, the associated
nuclear coupling is a conventional spin operator, $\Sigma_1^\prime$. 
\item By adopting an interaction having the form $\CO_1$ or $\CO_4$, one builds in
the assumption that detector rates depend on the $v^0$ moment of the galactic velocity
distribution.  This assumption is generally in error for operators other than $\CO_1$ and
$\CO_4$, even if the operator is one of those described in points 2 and 3 above, with
nuclear physics quite
similar to $\CO_1$ and $\CO_4$.  The rates for LO, NLO, NNLO,..., operators depend
on the $v^0$, $v^2$, $v^4$, ..., moments, respectively, of the WIMP
velocity distribution.  Consequent the distribution of events as a function of
recoil energy $E_R$
could be used to discriminate among classes of candidate interactions.
\end{enumerate}

\section{Nuclear Structure Input: Density Matrices}
\label{sec:sec5}
The Mathematica script described in Sec. \ref{sec:sec7} is designed to allow a
nuclear structure theorist to supply alternative descriptions of the nuclear physics, without requiring
him/her to have any detailed knowledge of the operator matrix elements that must
be calculated.   This is accomplished through the use of density matrices.

The dark matter particle scattering cross sections are expressed in terms of the single-reduced
 (in angular momentum) matrix elements of
one-body operators of definite angular momentum.  In the treatment so far we have
labeled the nuclear ground state by its angular momentum $j_N$, an exact quantum
number.  Here we add to that label the isospin quantum numbers $T,M_T$:  
isospin $T$ is an approximate but not exact quantum label, as isospin is broken
by the electromagnetic interactions among nucleons.  However, we employ that
label here because most shell-model calculations are isospin conserving, and 
thus most density matrices derived from such calculations employ $T$ as a 
quantum label.  We stress, however, that everything discussed below can be
trivially repeated without the assumption of $T$ as a nuclear state label:
 the density matrix would then be defined without this assumption.

The reduced (in angular momentum) many-body matrix element of an arbitrary one-body operator can be expressed
as a product of the one-body density matrix  and the single-particle matrix elements
of the one-body operator.  For our elastic case,
\begin{eqnarray}
\langle j_N; T M_T ||\sum_{i=1}^A \hat{O}_{J;\tau} (q \vec{x}_i) || j_N; T M_T \rangle &=& (-1)^{T-M_T} \left( \begin{array}{ccc}
T & \tau & T \\ -M_T & 0 & M_T \end{array} \right) \langle j_N;T~ \vdots \vdots \sum_{i=1}^A \hat{O}_{J;\tau}(q \vec{x}_i)  \vdots \vdots ~j_N; T \rangle \nonumber \\
\langle j_N;T~ \vdots \vdots \sum_{i=1}^A \hat{O}_{J;\tau}(q \vec{x}_i)  \vdots \vdots ~j_N; T \rangle&=&  \sum_{|\alpha|,|\beta|} \Psi_{|\alpha|,|\beta|} ^{J;\tau} ~\langle |\alpha|~ \vdots \vdots O_{J;\tau}(q \vec{x}) \vdots \vdots~ |\beta| \rangle.
\label{eq:densitymatrix}
\end{eqnarray} 
Here $\Psi_{|\alpha|,|\beta|} ^{J;\tau}$ is the one-body density matrix for the diagonal 
ground-state-to-ground-state transition; $|\alpha|$ represents the nonmagnetic
quantum numbers in the chosen single-particle basis (e.g., for a single-particle harmonic oscillator
state
$|\alpha \rangle = |n_\alpha(l_\alpha s_\alpha=1/2)j_\alpha m_{j_\alpha}; t_\alpha=1/2 m_{t_\alpha}  \rangle\equiv | |\alpha|; m_{j_\alpha} m_{t_\alpha} \rangle$,  with $n_\alpha$ the nodal quantum
number); $\vdots \vdots$ denotes a doubly reduced matrix element (in
spin and isospin); and the sums over $|\alpha|$ and $|\beta|$ extend over complete sets
of single-particle quantum numbers. The density matrix can be written in second
quantization as
\be
\Psi_{|\alpha|,|\beta|} ^{J;\tau} \equiv{1 \over \sqrt{ (2J+1)(2 \tau+1)}} ~\langle j_N;T~ \vdots \vdots
\left[c_{|\alpha|}^\dagger \otimes  \tilde{c}_{|\beta|} \right]_{J; \tau} \vdots \vdots~ j_N; T \rangle
\ee
where the single particle creation operator is  $c^\dagger_\alpha$ while
$\tilde{c}_\beta = (-1)^{j_\beta-m_{j_\beta}+1/2-m_{t_\beta}}  c_{|\beta|;- m_{j_\beta}, -m_{t_\beta}}$.  The phases yield a destruction operator  $\tilde{c}_\beta$ that transforms
as a spherical tensor in single-particle angular momentum and isospin.

Equation (\ref{eq:densitymatrix}) is an exact expression for
$\langle j_N; T M_T || \hat{O}_J;\tau || j_N; T M_T \rangle$.   When one invokes a nuclear model to calculate
a dark-matter response function, effectively one is employing some physics-motivated
prescription for intelligently  truncating the infinite sums over $|\alpha|,|\beta|$ 
in Eq. (\ref{eq:densitymatrix}) to some finite subset,
hopefully capturing most of the relevant low-momentum physics.

This expression factors a matrix element into a product of density matrix elements, which are
independent of any details of the operator $\hat{O}_{J;\tau}$ apart from its rank in 
angular momentum and isospin, and single-particle operator matrix elements.
The Mathematica script calculates the latter, assuming that the single-particle
basis for the Slater determinants is the harmonic oscillator.
Thus one can modify the script to use different nuclear physics input simply by
supplying alternative one-body density matrices for the nuclear targets of interest.  There
is no need to evaluate any operator matrix elements.
In making such substitutions (or better, in adding alternative density matrices, so that one
can begin to assess nuclear structure uncertainties), users should employ the single-particle conventions of \cite{donnellyhaxton}, for consistency with those employed in evaluating the single-particle
matrix elements of dark-matter operators.

The isospin matrix element in Eq. (\ref{eq:densitymatrix}) is easily performed, yielding
\begin{eqnarray}
\langle |\alpha|~ \vdots \vdots \hat{O}_J;\tau(q \vec{x}) \vdots \vdots~ |\beta| \rangle = \sqrt{2(2 \tau+1)}~ ~ \langle n_\alpha (l_\alpha 1/2)j_\alpha ~|| O_J ||~ n_\beta (l_\beta 1/2) j_\beta \rangle 
\label{eq:densitymatrixfinal}
\end{eqnarray}
where $O_J$ 
is the space-spin part of the operator.
For the harmonic oscillator basis used in the Mathematica script,  the single-particle reduced matrix element 
for $O_J = \{M_J$, $\Sigma^\prime_J$, $\Sigma^{\prime \prime}_J$, $\Delta_J$, $\tilde{\Phi}^\prime_J$, $\Phi^{\prime \prime}_J \}$ can  be evaluated algebraically,
\be
\langle n_\alpha (l_\alpha 1/2)j_\alpha || O_J(q \vec{x}) || n_\beta (l_\beta 1/2)j_\beta \rangle =
{1 \over \sqrt{4 \pi}} y^{(J-K)/2} e^{-y} p(y)
\ee
where $K=2$ for the normal parity 
($\pi=(-1)^J$) operators $M_J$, $\tilde{\Phi}_J^\prime$, and $\tilde{\Phi}_J^{\prime \prime}$ and $K=1$ for the abnormal parity ($\pi=(-1)^{J+1}$) operators
$\Delta$, $\Sigma^\prime$, and $\Sigma^{\prime \prime}$.  $y=(q b/2)^2$ where $b$ is the
oscillator parameter, and $p(y)$ is a finite polynomial in $y$.  Thus the
nuclear response functions $W$ of Eq. (\ref{eq:nucresponse}) are simple functions of $y$.

\begin{table}[htbp]
\begin{center}
\hspace*{-0.8cm}
\scalebox{0.90}{
\begin{tabular}{|c|c|c|c|c|}
\hline
\hline
& & & & \\
 $j$ & ${\CL}^j_\mathrm{int}$ & Nonrelativistic Reduction  &~~$ \displaystyle\sum_i c_i \CO_i$  &~~ P/T~~ \\[0.4cm]
\hline
& & & & \\
1&$ \bar{\chi} \chi \bar{N} N$ &$ 1_\chi 1_N $&$ \CO_1$ & E/E \\[0.3cm]
2&$ i \bar{\chi} \chi \bar{N} \gamma^5 N$ &$ i \displaystyle{{\vec{q} \over m_N} \cdot \vec{S}_N}$ & $\CO_{10}$ &O/O \\[0.3cm]
3&$ i \bar{\chi} \gamma^5 \chi \bar{N} N$ &$- i \displaystyle{{\vec{q} \over m_\chi} \cdot \vec{S}_\chi}$ & $-\displaystyle{{m_N \over m_\chi} \CO_{11}}$& O/O \\[0.3cm]
4&$ \bar{\chi} \gamma^5 \chi \bar{N} \gamma^5 N $&$- \displaystyle{{\vec{q} \over m_\chi} \cdot \vec{S}_\chi {\vec{q} \over m_N} \cdot \vec{S}_N}$  &$\displaystyle{- {m_N \over m_\chi} \CO_6}$ & E/E \\[0.3cm]
5&$\displaystyle{P^\mu \over m_\mathrm{M}} \bar{\chi} \chi \displaystyle{K_\mu \over m_\mathrm{M}} \bar{N} N $& 4$\displaystyle{{ m_\chi m_N\over m_\mathrm{M}^2}} 1_\chi 1 _N$&$4 \displaystyle{m_\chi m_N\over m_\mathrm{M}^2} \CO_1$ &E/E \\[0.3cm]
6&$\displaystyle{P^\mu \over m_\mathrm{M}} \bar{\chi} \chi \bar{N} i \sigma_{\mu \alpha} \displaystyle{q^\alpha \over m_\mathrm{M}} N$ &$-\displaystyle{m_\chi \over m_N} \displaystyle{\vec{q}^{\,2} \over m_\mathrm{M}^2} 1_\chi 1_N -4i \displaystyle{m_\chi \over m_\mathrm{M}} \vec{v}^\perp \cdot \left( \displaystyle{\vec{q} \over m_\mathrm{M}} \times \vec{S}_N \right)  $&$-\displaystyle{m_\chi \over m_N} \displaystyle{\vec{q}^{\,2} \over m_\mathrm{M}^2} \CO_1 
+4 \displaystyle{{m_\chi m_N \over m_\mathrm{M}^2}} \CO_3 $& E/E \\[0.3cm]
7&$\displaystyle{P^\mu \over m_\mathrm{M}} \bar{\chi} \chi \bar{N} \gamma_\mu \gamma^5 N$ & $-4 \displaystyle{m_\chi \over m_\mathrm{M}} \vec{v}^\perp \cdot \vec{S}_N$ & $-4 \displaystyle{m_\chi \over m_\mathrm{M}} \CO_7$  &O/E \\[0.3cm]
8& $i \displaystyle{P^\mu \over m_\mathrm{M}} \bar{\chi} \chi \displaystyle{K_\mu \over m_\mathrm{M}} \bar{N} \gamma^5 N$ & $4 i \displaystyle{{m_\chi \over m_\mathrm{M}}{\vec{q} \over m_\mathrm{M}} \cdot \vec{S}_N}$ & $4 \displaystyle{m_\chi m_N \over m_\mathrm{M}^2} \CO_{10}$ &O/O \\[0.3cm]
9& $\bar{\chi} i \sigma^{\mu\nu} \displaystyle{q_\nu \over m_\mathrm{M}} \chi \displaystyle{K_\mu \over m_\mathrm{M}} \bar{N} N $ &$\displaystyle{m_N \over m_\chi}\displaystyle{\vec{q}^{\,2} \over m_\mathrm{M}^2} 1_\chi 1_N +4i \displaystyle{m_N \over m_\mathrm{M}}\vec{v}^\perp \cdot \left( \displaystyle{\vec{q} \over m_\mathrm{M}} \times \vec{S}_\chi \right)$ & $\displaystyle{m_N \over m_\chi} \displaystyle{\vec{q}^{\,2} \over m_\mathrm{M}^2} \CO_1 -4  \displaystyle{m_N^2 \over m_\mathrm{M}^2}\CO_5$& E/E\\[0.3cm]
10& $\bar{\chi} i \sigma^{\mu\nu} \displaystyle{q_\nu \over m_\mathrm{M}} \chi \bar{N} i \sigma_{\mu\alpha} \displaystyle{q^\alpha \over m_\mathrm{M}} N$ &$ 4 \left( \displaystyle{\vec{q} \over m_\mathrm{M}} \times \vec{S}_\chi \right) \cdot \left( \displaystyle{\vec{q} \over m_\mathrm{M}} \times \vec{S}_N \right)$ & $4 \left( \displaystyle{\vec{q}^{\,2} \over m_\mathrm{M}^2} \CO_4 -\displaystyle{m_N^2 \over m_\mathrm{M}^2}\CO_6 \right) $&E/E \\[0.3cm]
11& $\bar{\chi} i \sigma^{\mu\nu} \displaystyle{q_\nu \over m_\mathrm{M}} \chi \bar{N} \gamma^\mu \gamma^5 N $& $-4 i  \left( \displaystyle{\vec{q} \over m_\mathrm{M}} \times \vec{S}_\chi \right) \cdot \vec{S}_N $ & $-4 \displaystyle{m_N \over m_\mathrm{M}} \CO_9$ & O/E\\[0.3cm]
12 & $i \bar{\chi} i \sigma^{\mu\nu} \displaystyle{q_\nu \over m_\mathrm{M}} \chi \displaystyle{K_\mu \over m_\mathrm{M}} \bar{N} \gamma^5 N$ &$ \left[i \displaystyle{\vec{q}^{\,2} \over m_\chi m_\mathrm{M}} -4 \vec{v}^\perp \cdot \left( \displaystyle{\vec{q} \over m_\mathrm{M}} \times \vec{S}_\chi \right)\right] \displaystyle{\vec{q} \over m_\mathrm{M}}  \cdot \vec{S}_N$ &$\displaystyle{m_N \over m_\chi}\displaystyle{\vec{q}^{\,2} \over  m_\mathrm{M}^2}  \CO_{10}+4\displaystyle{\vec{q}^{\,2} \over m_\mathrm{M}^2}   \CO_{12} + 4 \displaystyle{m_N^2 \over m_\mathrm{M}^2} \CO_{15}$ &O/O \\[0.3cm]
13 & $\bar{\chi} \gamma^\mu \gamma^5 \chi \displaystyle{K_\mu \over m_\mathrm{M}} \bar{N} N $&$4 \displaystyle{m_N \over m_\mathrm{M}}\vec{v}^\perp \cdot \vec{S}_\chi$ &$4 \displaystyle{m_N \over m_\mathrm{M}}\CO_8$ &O/E \\[0.3cm]
14 &$\bar{\chi} \gamma^\mu \gamma^5 \chi \bar{N} i \sigma_{\mu\alpha} \displaystyle{q^\alpha \over m_\mathrm{M}} N $&$-4i \vec{S}_\chi \cdot \left( \displaystyle{\vec{q} \over m_\mathrm{M}} \times \vec{S}_N \right) $ & $4 \displaystyle{m_N \over m_\mathrm{M}}\CO_9$ &O/E \\[0.3cm]
15 & $\bar{\chi} \gamma^\mu \gamma^5 \chi \bar{N} \gamma^\mu \gamma^5 N$ &$-4 \vec{S}_\chi \cdot \vec{S}_N$ & $-4 \CO_4$ &E/E  \\[0.3cm]
16 & $i \bar{\chi} \gamma^\mu \gamma^5 \chi \displaystyle{K^\mu \over m_\mathrm{M}} \bar{N} \gamma^5 N$ & $4i \vec{v}^\perp \cdot \vec{S}_\chi \displaystyle{\vec{q} \over m_\mathrm{M}} \cdot \vec{S}_N$  &$ 4 \displaystyle{m_N \over m_\mathrm{M}}\CO_{13}$& E/O  \\[0.3cm]
17 & $i \displaystyle{P^\mu \over m_\mathrm{M}} \bar{\chi} \gamma^5 \chi \displaystyle{K_\mu \over m_\mathrm{M}} \bar{N} N $& $-4i \displaystyle{m_N \over m_\mathrm{M}}\displaystyle{\vec{q} \over m_\mathrm{M}} \cdot \vec{S}_\chi $& $-4 \displaystyle{m_N^2 \over m_\mathrm{M}^2} \CO_{11}$ &O/O  \\[0.3cm]
18 & $i \displaystyle{P^\mu \over m_\mathrm{M}} \bar{\chi} \gamma^5 \chi \bar{N} i \sigma_{\mu \alpha} \displaystyle{q^\alpha \over m_\mathrm{M}} N $ & $\displaystyle{\vec{q} \over m_\mathrm{M}} \cdot \vec{S}_\chi \left[ i\displaystyle{\vec{q}^{\,2} \over m_N m_\mathrm{M}} -4 \vec{v}^\perp \cdot \left( \displaystyle{\vec{q} \over m_\mathrm{M}} \times \vec{S}_N \right) \right]$ & $ \displaystyle{\vec{q}^{\,2} \over m_\mathrm{M}^2} \CO_{11} +4 \displaystyle{m_N^2 \over m_\mathrm{M}^2} \CO_{15} $ &O/O  \\[0.3cm]
19& $i \displaystyle{P^\mu \over m_\mathrm{M}} \bar{\chi} \gamma^5 \chi \bar{N} \gamma_\mu \gamma^5 N $& $4i \displaystyle{\vec{q} \over m_\mathrm{M}} \cdot \vec{S}_\chi \vec{v}_\perp \cdot \vec{S}_N$& $4  \displaystyle{m_N \over m_\mathrm{M}}\CO_{14}$ & E/O \\[0.3cm]
20 & $\displaystyle{P^\mu \over m_\mathrm{M}} \bar{\chi} \gamma^5 \chi \displaystyle{K_\mu \over m_\mathrm{M}} \bar{N} \gamma^5 N$   &$-4 \displaystyle{\vec{q} \over m_\mathrm{M}} \cdot \vec{S}_\chi \displaystyle{\vec{q} \over m_\mathrm{M}} \cdot \vec{S}_N$ &$-4 \displaystyle{m_N^2 \over m_\mathrm{M}^2} \CO_6$ & E/E \\
 & &  & & \\
 \hline
\end{tabular}}
\caption{ The Lagrangian densities $\CL^j_\mathrm{int}$, the operators 
obtained after nonrelativistic reduction that would be used between Pauli spinors to generate
the invariant amplitude, the 
corresponding effective interactions in terms of the EFT operators, and the transformation properties 
of the interactions (even E or odd O) under parity and time reversal.  Bjorken and Drell spinor and
gamma matrix conventions are used.  The scale $m_\mathrm{M}$, which usually would be
known from the context of the theory, can be put into the Mathematica script, or set to its default
value, $m_v$ of Sec. \ref{sec:sec2pt2}.  See Sec. \ref{sec:sec6pt2} for further discussion.}
 \label{table:LWL}
\end{center}
\end{table}

\section{Particle Theory Input: Nonrelativistic Matching}
\label{sec:sec6}
In most cases a theorist interested in a given ultraviolet theory of dark matter will derive 
a relativistic WIMP-nucleon interaction $\mathcal{L}_\mathrm{int}$.  The Mathematica script is set up to allow one
to 1) input the coefficients $c_i$ of the nonrelativistic Galilean-invariant operators $\CO_i$
directly (the choice one would likely make in experimental analysis);  or
alternatively 2) input the coefficients $d_j$ of a set of such covariant interactions 
amplitudes $\mathcal{L}^j_\mathrm{int}$, which are then reduced to the form $\sum_i c_i \CO_i$.

\subsection{The $\mathcal{L}^j_{int}$}
 In Sec. \ref{sec:sec2} we discussed 
two simple examples of 2), the spin-independent
and spin-dependent interactions $\mathcal{L}^\mathrm{SI}_\mathrm{int}$ and
$\mathcal{L}^\mathrm{SD}_\mathrm{int}$.  The invariant amplitudes obtained from these $\mathcal{L}$
were reduced to their nonrelativistic forms, yielding matrix elements between Pauli spinors.
The nonrelativistic operators between the Pauli spinors were identified as $\CO_1$ and
$\CO_4$.  Here we repeat the process for large set of $\mathcal{L}_{int}^j$ 
listed in Table \ref{table:LWL}.  Unlike the simple cases discussed in Sec. \ref{sec:sec2},
the $\mathcal{L}_{int}^j$ do not always map onto single nonrelativistic operators $\CO_j$.
Instead the result is frequently
\be
\mathcal{L}^j_\mathrm{int} ~ \rightarrow ~ \sum_i c_i(j) \CO_i.
\ee
where several $c_i(j)$ are nonzero.

The interactions of Table \ref{table:LWL}, coded into our Mathematic script, describe the interactions of spin-1/2 WIMPS with nucleons.  (More general interactions could be considered,
of course.)  Four-momentum definitions follow our three-momentum conventions: the incoming (outgoing) four-momentum of the dark matter particle $\chi$ is $p^\mu$ ($p^{\prime \, \mu}$);
the incoming (outgoing) four-momentum of the nucleon $N$ is $k^\mu$ ($k^{\prime \, \mu}$); and
the momentum transfer $q^\mu = p^{\prime \, \mu}-p^\mu=k^\mu - k^{\prime \, \mu}$.
We also define $P^\mu=p^\mu+ p^{\prime \, \mu}$ and
$K^\mu=k^\mu + k^{\prime \, \mu}$.  The relative velocity operator of Eq. (\ref{eq:vperp})
can be written in term of these variables as
\be
\vec{v}^\perp \equiv  {1 \over 2} \left( \vec{v}_{\chi,\mathrm{in}}+\vec{v}_{\chi,\mathrm{out}} - \vec{v}_{N,\mathrm{in}} - \vec{v}_{N,\mathrm{out}} \right)  ={1 \over 2} \left( {\vec{P} \over m_\chi} - {\vec{K} \over m_N} \right).
 \ee
The relativistic WIMP-nucleon interactions are constructed as bilinear  WIMP-nucleon products
of the available scalar  ($\bar{\chi} \chi$, $\bar{\chi}\gamma^5 \chi$) and four-vector ($\bar{\chi} P^\mu \chi$,
$\bar{\chi} P^\mu \gamma^5 \chi$, $\bar{\chi} i  \sigma^{\mu \nu} q_\nu \chi$, and $\bar{\chi} \gamma^\mu \gamma^5 \chi$) amplitudes.  Thus there are
$2^2 + 4^2=20$ combinations \cite{nonrelEFT}.  The nonrelativistic operators obtained
after nonrelativistic reduction are listed in Table \ref{table:LWL}, along with the corresponding
expansions in terms of our EFT operators, the $\CO_i$.  The Table also
gives transformation properties of the interactions under parity and time-reversal.  Note that
all interactions reduce in leading order to combinations of our fifteen $\CO_i$, and all of the 
$\CO_i$ appear in the Table.  Thus they are the minimal set of nonrelativistic interactions needed
to represent the listed set of 20 $\mathcal{L}^j_{int}.$

\subsection{Units: Inputing the $d_j$s into the Mathematic Script}
\label{sec:sec6pt2}
The Mathematica script allows one to specific an interaction corresponding to an arbitrary 
sum over the terms in Table \ref{table:LWL}.  It then calculates the corresponding operator,
expressing it in terms of the $\CO_i$, that one uses between Paul spinors to calculate
the WIMP-nucleon (Galilean) invariant amplitude. Thus
\be
\sum_j d_j \mathcal{L}_{int}^j \rightarrow \sum_i c_i \CO_i
\ee
where the $c_i$ are functions of the $\{d_j\}$.  The $\mathcal{L}_{int}^j$ of the Table all have
the same dimension, as the dimensionless quantities $P_\mu/m_M$, $K_\mu/m_M$ and
$q_\mu/m_M$ were used in the operator construction.  Here $m_M$ is a mass scale that will
generally be known, given a model context.  The $d_j$, like the $c_j$, have dimensions 1/mass$^2$.
Hence the conversion of a $d_j$ to a linear combination of the $c_i$ involves an expression
with only dimensionless coefficients.

In analogy with input for the $c_j$, the $d_j$ are input in units of $1/m_M^2$.  That is
\be
\mathrm{input~}d_j=1 \Rightarrow d_j=1/m_M^2.
\label{eq:dconversion}
\ee
The user is queried about the desired input value for $m_M$.  If no value is specified, the
script defaults to the choice $m_M=m_v=246.2$ GeV, weak interaction strength.  
Note that the value of $m_M$ is used both in Eq. (\ref{eq:dconversion}), converting input
numbers into appropriately dimensioned couplings $d_j$, as well as in defining the momentum-dependent
operators appearing in the Table.

\section{The Mathematica Script: Documentation}
\label{sec:sec7}

The formalism presented in this paper, with its factorization cross sections into products of 
WIMP and nuclear responses, is the basis for the Mathematica script presented here.  The
script was constructed so that experimental groups would be able to conduct model
independent analyses of their experiments using the EFT framework.  We have integrated
the particle and nuclear physics in ways that should make the code useful to nuclear
structure and particle theorists as well, as described in previous sections.

In this section, which also serves as a \verb|readme| file for the program, we discuss the usage of the program itself.

\subsection{Initialization}

Our Mathematica package, along with all of the associated documentation, can be found at  \\
{\color{blue} \url{http://www.ocf.berkeley.edu/~nanand/software/dmformfactor/}}. To initialize the package, either put \verb|dmformfactor.m| in your directory for Mathematica packages and run

\begin{Verbatim}[frame=single,
       framerule=0.2mm,framesep=3mm,fillcolor=\color{Gray}]
<<`dmformfactor
\end{Verbatim}

or initialize the package file itself from its source directory. For example

\begin{Verbatim}[frame=single,
       framerule=0.2mm,framesep=3mm,fillcolor=\color{Gray}]
<<"/Users/me/myfiles/dmformfactor.m"
\end{Verbatim}

\subsection{Summary of Functions}

In order to compute the  WIMP response functions $R_i^{\tau \tau^\prime}(\vec{v}_T^{\perp 2}, {\vec{q}^{\,2} \over m_N^2})$, the user
 must first call functions setting the dark matter mass and spin as well as the coefficients of the effective Lagrangian.  In order to compute the nuclear response functions $W_i((qb/2)^2)$, the user must 
 specify the $Z$ and $A$ of the isotope.  The density matrices and the oscillator parameter
 $b$ needed in the calculation of the $W_i$ are set internally in the script, though there are options to override
 the internal values.  The nuclear ground state spin and isospin (the script assumes 
 exact isospin, consistent with an input
 density matrix that is doubly reduced - see text) are also set internally, once $Z$ and $A$ are input.

\begin{itemize}

\item \verb|SetJChi| and \verb|SetMChi|: These set the dark matter spin and mass, respectively.   Simply call:

\begin{Verbatim}[frame=single,
       framerule=0.2mm,framesep=3mm,fillcolor=\color{Gray}]
SetJChi[j]
\end{Verbatim}

and 

\begin{Verbatim}[frame=single,
       framerule=0.2mm,framesep=3mm,fillcolor=\color{Gray}]
SetMChi[m]
\end{Verbatim}

to set the dark matter spin to \verb|j| and the dark matter mass to \verb|m|. The unit \verb|GeV| is recognized by the script; for example, calling \verb|SetMChi[10 GeV]| sets the dark matter mass to 10 GeV.

\item \verb|SetIsotope[Z,A,bFM, filename]|

 This sets the nuclear physics input, including the charge \verb|Z| and atomic number \verb|A| of the isotope, the file for the density matrices that the user wants to use, and the oscillator parameter $b[\textrm{fm}]$ (that is, $b$ in femtometers).  If the users elects to use the default density matrices (which are 
available for ${}^{19}$F, ${}^{23}$Na, ${}^{70}$Ge, ${}^{72}$Ge, ${}^{73}$Ge, ${}^{74}$Ge, ${}^{76}$Ge, ${}^{127}$I, ${}^{128}$Xe, ${}^{129}$Xe, ${}^{130}$Xe, ${}^{131}$Xe, ${}^{132}$Xe, ${}^{134}$Xe, and ${}^{136}$Xe), then simple take \verb|filename| to be ``default'' (note that one must still specify the correct \verb|Z| and \verb|A| for the isotope of interest.  Otherwise, users must provide their own density matrix file, to be read in by the program.  Similarly, entering ``default" for $b$ will employ
the approximate formula $b[\textrm{fm}]= \sqrt{41.467/(45A^{-1/3}-25A^{-2/3})}$.  To use another value of
$b[\textrm{fm}]$, enter a numerical value.  The nuclear mass is set to $A m_N$.  

\item \verb|SetCoeffsNonrel[i,value,isospin]|

This sets the coefficients $c_i$ of the EFT operators
$\CO_i$.  The script allows the user to set
values for $\{c_1,c_3,c_4, ..., c_{15} \}$; note that $c_2$ is excluded, for reasons discussed
in the text. We have chosen a normalization such that the coefficients $c_i$ all have dimensions $(\textrm{Energy})^{-2}$;\footnote{Note that this convention for the $c_i$'s differs from that in \cite{nonrelEFT}.} to compensate for this, the dimensionless user input for \verb|value| is multiplied by $m_V^{-2}$, with  $m_V \equiv 246.2$ GeV.

The coefficients carry an isospin index $\alpha$ that can be specified in one of two ways, as a coupling to protons and neutrons, $\{c_i^p,c_i^n\}$, in which case the associated operator is
\be
                                 \left[ c_i^p {1 + \tau_3 \over 2} + c_i^n {1 - \tau_3 \over 2} \right] \CO_i
\ee 
or as a coupling to isospin, $\{c_i^0,c_i^1\}$, where the associated operator is
\be
                                 \left[ c_i^0  + c_i^1\tau_3  \right] \CO_i.
\ee
For the former, the input should be ``n'' for neutrons and ``p'' for protons. For example:

\begin{Verbatim}[frame=single,
       framerule=0.2mm,framesep=3mm,fillcolor=\color{Gray}]
SetCoeffsNonrel[4,12.3, "p"]
\end{Verbatim}  

whereas for the latter it should be 0 for isoscalar and 1 for isovector.  All coefficients are set to 0 by default when the package is initialized.  \verb|SetCoeffsNonrel| will change only the coefficient specified, and will leave all other coefficients unchanged.  So, for example, if one initializes the package and calls \verb|SetCoeffsNonrel[4,12.3, 0]|, then $c_4^{p}$ and $c_4^{n}$ will both be 6.15, with all other coefficients vanishing.  If one then calls \verb|SetCoeffsNonrel[4,3.3,``p'']|, then $c_4^{p}$ will be set to 3.3, but $c_4^{n}$ will not change and will still be 6.15. Thus by making two calls, an arbitrary combination 
of $\{c_4^{p},c_4^{n} \}$ or equivalently $\{c_4^0, c_4^1\}$ can be set.

\item \verb|SetCoeffsRel[i,value,isospin]|

These functions are similar to \verb|SetCoeffsNonrel|, except that they set the 
coefficients 
$d_j$ of the 20 covariant
interactions $\CL_\mathrm{int}^j$ defined in Table \ref{table:LWL}.
The coefficients $d_j$ are dimensionless, by inserting appropriate powers of the user-defined scale
$m_M$, set by the user function \verb|SetMM|.  This scale is set by default to be $m_M=m_V \equiv 246.2$ GeV.
We adopt a  convention
where the spinors in  $\CL_\mathrm{int}^j$  are defined as normalized to unity: with this convention
a nonrelativistic reduction of the $\CL_\mathrm{int}^j$ in the second column of Table \ref{table:LWL}
would give the results in the fourth column.  
[As noted in the paper, we use a spinor normalization of $2m$ in our derivations, but extract
the factor of $4m_\chi m_N$ in order to maintain the definition above.]

\verb|SetCoeffsNonrel| and \verb|SetCoeffsRel| cannot be used together.  By default, the package assumes you will use \verb|SetCoeffsNonrel|.  The first time the user calls \verb|SetCoeffsRel|, the package will first reset all coefficients back to zero before calling \verb|SetCoeffsRel|, after which point it will act normally.  A subsequent call to \verb|SetCoeffsNonrel| will similarly first reset all coefficients back to zero and then revert to non-relativistic mode.  

Since the relativistic operators implicitly assume spin-1/2 WIMPs, any call to \verb|SetCoeffsRel| automatically sets $j_\chi = 1/2$.

\item \verb|SetMM[mM]|

Set the fiducial scale $m_M$ for the relativistic coefficients $d_i$.

\item \verb|ZeroCoeffs[]|

Calling \verb|ZeroCoeffs[]| simply resets all operators coefficients to zero.

 \item \verb|ResponseNuclear[y,i,tau,tau2]|

This function prints out any of the eight nuclear response functions $W_i^{\tau \tau_2}(y)$.
This involves a folding of the single-particle matrix elements
with the density matrices.  The results are printed as analytic functions
in the dimensionless variable $y=(q b/2)^2$.  The $i$ run from 1 to 8, according to 
1) $W_M$, 2) $W_{\Sigma''}$, 3) $W_{\Sigma'}$, 4) $W_{\Phi''}$, 5) $W_{\tilde{\Phi}'}$, 6) $W_{\Delta}$, 7) $W_{M \Phi''}$, and 8) $W_{\Sigma' \Delta}$.

 \item \verb|TransitionProbability[v,q(,IfRel)]|
 
This is the main user function.  It first prints out the Lagrangian that is being used.

Second, it folds the $W_i^{\tau \tau^\prime}(y)$ and $R_i^{\tau \tau^\prime}(\vec{v}_T^{\perp 2}, {\vec{q}^{\,2} \over m_N^2 })$ to form 
\be
P_{\rm tot} = {1 \over 2j_\chi + 1} {1 \over 2j_N + 1} \sum_\mathrm{spins} |\CM|^2_\mathrm{nucleus-HO/EFT},
\label{eq:transitionprob}
\ee

It then evaluates the transition probability for the numerical values of $b$ and $m_N$.  As $b$
is in fm, the substitution is $y=(qb/(2 \hbar c))^2 \sim (q b/2(0.197\mathrm{ Gev~ fm}))^2$.  As $m_N$ is input in GeV,
this evaluates Eq. (\ref{eq:HamHO}) as a function \verb|TransitionProbability[vsq,q]| where \verb|q| is in GeV.
This function can be printed out or plotted numerically.

The conventional relativistic normalization of the amplitude differs from the non-relativistic normalization by a factor
of $1/(4 m_\chi m_T)$.  Since the conventional relativistic normalization is commonly used and produces a dimensionless value for $|\CM|^2$, we also provide an optional argument \verb|IfRel|, which if set to \verb|True| will output
(\ref{eq:transitionprob}) with the relativistic normalization convention (that is, it will multiply by $(4 m_\chi m_T)^2$ to produce a dimensionless transition probability).  By default, it is set to \verb|False|.

\item \verb|DiffCrossSection[ERkeV,v]|

From the transition probability $P_{\rm tot}$, one can immediately obtain the differential cross section per recoil energy:
\be
\frac{d \sigma}{d E_R} &=& \frac{m_T}{2\pi v^2} P_{\rm tot}.
\ee
The function \verb|DiffCrossSection[ERkeV,v]| takes as arguments the recoil energy in units of keV and the velocity of the incoming DM particle in the lab frame.  It first prints out the Lagrangian being used, and then outputs the differential cross-section $\frac{d \sigma}{d E_R}$. 

\item \verb|ApproxTotalCrossSection[v]|

From the differential cross-section $\frac{d \sigma}{d E_R}$, one can also obtain the total cross-section as a function of $v$ by integrating over recoil energies.  In general, this depends on energy thresholds and, written in closed form, is a complicated analytic function due to the exponential damping factor $e^{-2y}$ in the response functions, so for precise values it is simplest to do the energy integration numerically.  However, for approximate results we can consider the limit of small nuclear harmonic oscillator parameter $b$, in which case the exponential factor $e^{-2y}$ can be neglected.  For fixed $v$, the integration over $E_R$ from zero up to the kinematic threshold $E_{R, \rm max} = 2 \frac{\mu_T^2 v^2}{m_T}$ can be performed analytically.  The function \verb|ApproxTotalCrossSection[v]| takes as argument the velocity $v$ of the incoming DM particle in the lab frame and, after printing out the Lagrangian being used, outputs this approximate total cross-section $\sigma(v)$.

\item \verb|EventRate[|$N_T$,$\rho_\chi$,$q$,$v_e$,$v_0$(,$v_{\rm esc}$)\verb|]|

One can determine the total detector event rate (per unit time per unit detector mass per unit recoil energy) in terms of the transition probability $P_{\rm tot}$.  One simply multiplies $P_{\rm tot}$ by the appropriate prefactor and integrates over the halo velocity distribution, as follows:
\be
\frac{d R_D}{d E_R} = N_T \frac{\rho_\chi m_T}{2 \pi m_\chi } \left\< \frac{1}{v} P_{\rm tot}(v^2, q^2) \right\>
\ee
Here, $\< \dots \>$ indicates averaging over the halo velocity distribution.  $N_T$ is the number of target nuclei per detector mass, $\rho_\chi$ is the local dark matter density, $m_\chi$ is the dark matter mass, and $m_N$ is the nucleon mass.  In general, the halo average integral should include a lower-bound on the magnitude of the velocity at $v_{\rm min}$, which is $v_{\rm min} = \frac{q}{2\mu_T}$ for elastic scattering:
\be
\< h(q,\vec{v}) \> &\equiv& \int_{v_{\rm min}(q)}^\infty v^2 dv \int d^2 \Omega f_v(\vec{v} + \vec{v}_e) h(q,\vec{v}) .
\ee
The vector $\vec{v}_e$ is the Earth's velocity in the galactic rest frame. 
 While there has been much work recently on understanding theoretical constraints on the halo distribution from N-body simulations and from general considerations of dynamics, little is known by direct observation and there are still large uncertainties.  A very simple approximation that suffices for general considerations is to take a Maxwell-Boltzmann distribution:
\be
f_v(\vec{v}) &=& \frac{1}{\pi^{3/2} v_0^3} e^{-v^2/v_0^2},
\ee
where $v_0$ is roughly 220 km/s, about the rms velocity of the visible matter distribution (though N-body simulations suggest that the dark matter distribution may be shallower, and a larger $v_0$ may be more appropriate). The function EventRate[q,b,v$_e$,v$_0$] evaluates the event rate $\frac{d R_D}{d E_R}$ assuming this Maxwell-Boltzmann distribution as default. A cut-off Maxwell-Boltzmann distribution is also implemented as an option, in which case
\be
f_v(\vec{v}) &\propto& \left(e^{-v^2/v_0^2} - e^{v_{\rm esc}^2/v_0^2}\right) \Theta(v_{\rm esc}^2 - \vec{v}^2)
\ee
where $v_{\rm esc}$ is the escape velocity, and the subtraction above is included to make the distribution shut down smoothly. In this case, $v_{\rm esc}$ should be included as an optional argument to EventRate; if it is not included, it is set to a default value of $12 v_0$ (which is essentially $v_{\rm esc} = \infty$).  

\item \verb|SetHALO[halo]|

This sets the halo distribution used.  The variable $halo$ can be set either to ``MB'', in which case the Maxwell-Boltzmann distribution is used, or ``MBcutoff'', in which case the cut-off Maxwell-Boltzmann distribution is used.  It is set to ``MB'' by default.  

\item \verb|SetHelm[UseHelm]|

Calling \verb|SetHelm[True]|  sets the structure function for the density operator $M_J$ to be given by the Helm form factor, rather than by the structure function obtained from the density matrix.  \verb|SetHelm[False]| implements the structure function based on the density matrix, which is the default setting.

\end{itemize}

\subsection{Examples}

A full example for the transition probability would look like the following:

\begin{Verbatim}[frame=single,
       framerule=0.2mm,framesep=3mm,fillcolor=\color{Gray}]
<< "/Users/me/mypackages/dmformfactor.m";
SetJChi[1/2]
SetMChi[50 GeV]
F19filename="default";
bFM="default";
SetIsotope[9, 19, bFM, F19filename]
SetCoeffsNonrel[3, 3.1, "p"]
TransitionProbability[v,qGeV]
TransitionProbability[v,qGeV,True]
\end{Verbatim}

To additionally calculate the event rate $\frac{d R_D}{d E_R}$ in a Maxwell-Boltzmann halo velocity distribution, one can call

\begin{Verbatim}[frame=single,
       framerule=0.2mm,framesep=3mm,fillcolor=\color{Gray}]
mNucleon=0.938 GeV;
NT=1/(19 mNucleon);
Centimeter=(10^13 Femtometer);
rhoDM=0.3 GeV/Centimeter^3;
ve=232 KilometerPerSecond;
v0=220 KilometerPerSecond;
EventRate[NT,rhoDM,qGeV,ve,v0]
\end{Verbatim}

For a cut-off Maxwell-Boltzmann halo, an escape velocity must also be specified:

\begin{Verbatim}[frame=single,
       framerule=0.2mm,framesep=3mm,fillcolor=\color{Gray}]
mNucleon=0.938 GeV;
NT=1/(19 mNucleon);
Centimeter=(10^13 Femtometer);
rhoDM=0.3 GeV/Centimeter^3;
ve=232 KilometerPerSecond;
v0=220 KilometerPerSecond;
vesc=550 KilometerPerSecond;
SetHalo["MBcutoff"];
EventRate[NT,rhoDM,qGeV,ve,v0,vesc]
\end{Verbatim}

Finally, to get a quick estimate of the experimental bound from the 225 live day run of XENON100, one can use the standard spin-independent isoscalar interaction for a generic isotope of xenon, taking xenon-131 for instance.  Taking into account efficiencies, the total effective exposure is approximately 2500 kg days. A relativistic operator coefficient of $2 f_p/{\rm GeV}^2$ with $f_p = 4\cdot 10^{-9}$ predicts only a couple of events, and so should be close to the upper limit of their allowed cross-section:

\begin{Verbatim}[frame=single,
       framerule=0.2mm,framesep=3mm,fillcolor=\color{Gray}]
mNucleon=0.938 GeV;
NT=1/(131 mNucleon);
Centimeter=(10^13 Femtometer);
rhoDM=0.3 GeV/Centimeter^3;
SetMChi[150 GeV]
ve=232 KilometerPerSecond;
v0=220 KilometerPerSecond;
vesc=550 KilometerPerSecond;
SetHALO["MBcutoff"];
Xe131filename="default";
bFM="default";
SetIsotope[54, 131, bFM, Xe131filename]
SetCoeffsRel[1,2fp,0]
myrate[qGeV_]=(2500 KilogramDay) EventRate[NT,rhoDM,qGeV,ve,v0,vesc];
fp=2.4*10^(-4);
NIntegrate[myrate[qGeV] GeV*(qGeV GeV/(131 mNucleon)),{qGeV,0,10}]
\end{Verbatim}

The final line of output should be 2.06 for the value of the integral, which gives the predicted number of events.  The factor $\frac{q}{131 m_N}= \frac{q}{m_T}$ inside the integral is from the change of variables from $d E_R$ to $dq$, since $E_R = q^2/2m_T$.  In this example, the WIMP is sufficiently heavy that the exact low-energy threshold changes the prediction by less than a factor of two, so to get a rough estimate we have just integrated down to zero energy.  Finally, we can look what nucleon scattering cross-section corresponds to $f_p = 2.4 \cdot 10^{-4}$:
\be
\sigma_p &=& \frac{(4 m_N m_T f_p/m_V^2)^2}{16 \pi (m_N + m_T)^2} = 1.7 \cdot 10^{-45} {\rm cm}^2
\ee
which agrees to within a factor of a few  with the published upper bound on $\sigma_p$ from the XENON100 collaboration \cite{xenon}.  A more accurate calculation of the bound would include, among other corrections, the exact energy thresholds in the integral over momentum transfer, an average over the year as the earth's velocity changes, a sum over different isotopes according to their natural abundance, and a more precise treatment of energy-dependent efficiencies.

\subsection{Density Matrix Syntax}

 If one calls \verb|SetIsotope[Z,A, filename]| with a custom density matrix, the input density matrix file must contain the reduced density matrix elements $\Psi^{J,T}(|\alpha|, |\beta|)$ to be used. The in and out states $|\alpha|$ and $|\beta|$ should be specified by their principle quantum number $N$ and their total angular momentum $j$. See \cite{sevenops} for more details. The format of the file for each projection onto operators of spin $J$ and isospin $J$ should be as follows:

\vspace{0.8cm}

\verb#  ONE-BODY DENSITY MATRIX  # \hspace{1.5 cm} $\cdots$ \hspace{1.5cm} \verb#2J0= #$2J$, \quad $\cdots$ \quad $2T$

\qquad $\cdots$ $N^1_{\rm in}$ \qquad $2j^1_{\rm in}$ \qquad $N^1_{\rm out}$ \qquad $2j^1_{\rm out}$ \qquad $\Psi^{J,T}\left( \{ N^1_{\rm in}, j^1_{\rm in} \}; \{ N^1_{\rm out}, j^1_{\rm out} \} \right)$

\hspace{2.5cm} \vdots \hspace{2.5cm} \vdots \hspace{2.5cm} \vdots

\qquad $\cdots$ $N^n_{\rm in}$ \qquad $2j^n_{\rm in}$ \qquad $N^n_{\rm out}$ \qquad $2j^n_{\rm out}$ \qquad $\Psi^{J,T}\left( \{ N^n_{\rm in}, j^n_{\rm in} \}; \{ N^n_{\rm out}, j^n_{\rm out} \} \right)$

\vspace{0.8cm}

Dots ``$\cdots$'' indicate places where the code will simply ignore what appears there - the routines reading in the input are searching for regular expressions that match the above syntax.  Consequently, additional lines in the file that are not of the above form will also be ignored.
This is probably clearest to follow by seeing an explicit example. For instance, the density matrix for ${}^{19}$F is:

\begin{Verbatim}[frame=single,
       framerule=0.2mm,framesep=3mm,fillcolor=\color{Gray}]
	  INITIAL STATE CHARGE CONJ SYM =  0  TIME REVERSAL SYM =  0

	  FINAL STATE CHARGE CONJ SYM =  0  TIME REVERSAL SYM =  0
	        -23.88003     -23.88003

  ONE-BODY DENSITY MATRIX FOR 2JF =  1 2TF =  1 2JI =  1 2TI =  1 2JO =  0  TO = , 0
	       NBRA   2*JBRA   NKET   2*JKET     VALUE
	        0       1       0       1        4.00000000
	        1       1       1       1        4.00000000
	        1       3       1       3        5.65685425
	        2       1       2       1        1.22525930
	        2       3       2       3        0.20366116
	        2       5       2       5        0.85835832

  ONE-BODY DENSITY MATRIX FOR 2JF =  1 2TF =  1 2JI =  1 2TI =  1 2JO =  0  TO = , 2
	       NBRA   2*JBRA   NKET   2*JKET     VALUE
	        2       1       2       1        0.36984837
	        2       3       2       3        0.04794379
	        2       5       2       5        0.32467225

  ONE-BODY DENSITY MATRIX FOR 2JF =  1 2TF =  1 2JI =  1 2TI =  1 2JO =  2  TO = , 0
	       NBRA   2*JBRA   NKET   2*JKET     VALUE
	        2       1       2       1        0.44514263
	        2       3       2       1       -0.01197751
	        2       1       2       3        0.01197751
	        2       3       2       3       -0.05428837
	        2       5       2       3       -0.12172578
	        2       3       2       5        0.12172578
	        2       5       2       5        0.12280637

  ONE-BODY DENSITY MATRIX FOR 2JF =  1 2TF =  1 2JI =  1 2TI =  1 2JO =  2  TO = , 2
	       NBRA   2*JBRA   NKET   2*JKET     VALUE
	        2       1       2       1       -0.40780345
	        2       3       2       1       -0.01278520
	        2       1       2       3        0.01278520
	        2       3       2       3        0.01209672
	        2       5       2       3        0.10547489
	        2       3       2       5       -0.10547489
	        2       5       2       5       -0.24110544
\end{Verbatim}

Example density matrix file shown for ${}^{19}$F. The density matrices for  ${}^{19}$F, ${}^{23}$Na, ${}^{70}$Ge, ${}^{72}$Ge,${}^{73}$Ge,${}^{74}$Ge,${}^{76}$Ge, ${}^{127}$I, ${}^{128}$Xe,${}^{129}$Xe,${}^{130}$Xe,${}^{131}$Xe,${}^{132}$Xe,${}^{134}$Xe, and ${}^{136}$Xe are already built into the program and no external file is needed.

\section*{Acknowledgments} 

We would like to acknowledge useful conversations and feedback on the program and draft from Spencer Chang, Tim Cohen, Eugenio Del Nobile, Paddy Fox, Ami Katz, and Tim Tait. NA thanks the UC MultiCampus Research Program `Dark Matter Search Initiative'
for support.  ALF was partially supported by ERC grant BSMOXFORD no. 228169. WH is supported by the US Department of Energy under contracts DE-SC00046548 and DE-AC02-98CH10886.

\clearpage
\appendix

\section{Appendix: Some Details of the Response Function Derivation}
\label{sec:nuclresp}
The algebraic techniques that lead to Eq. (\ref{eq:Ham}) are commonly used in treatments of
semi-leptonic weak interactions.  We will briefly outline the steps, after first taking note of
certain simplifications that are made in the many-body theory to obtain the relatively
tractable form of Eq. (\ref{eq:Ham}).

\subsection{Treatment of the velocity operator}
We take as our  WIMP-nucleus interaction  the
sum over the one-body interactions of the WIMP with the individual nucleons in the nucleus.
While this is the usual starting point for treatments of electroweak nuclear reactions, 
it is an assumption.  The nucleon is a composite object held together by the exchange of
various mesons, which clearly can have their own interactions with the WIMP.  There has been
some work on the possible size of two-body corrections to WIMP-nucleus interactions
\cite{cirigliano,schwenk}.  Our feeling at this point is that the uncertainty of
the WIMP interaction with nucleons, as embodied in our fourteen coefficients $c_i$, is
currently so great that the one-body approximation is appropriate.  This sentiment would 
change were dark matter interactions discovered, making a detailed understanding
WIMP-matter interactions important.

Given the assumption of a one-body interaction, we noted that the Galilean invariance then leads to
the replacement
\begin{eqnarray}
\label{eq:separation}
\vec{v}^\perp &\rightarrow& \left\{ {1 \over 2} \left( \vec{v}_{\chi,\mathrm{in}} + \vec{v}_{\chi,\mathrm{out}}-
\vec{v}_{N,\mathrm{in}}(i)-\vec{v}_{N,\mathrm{out}}(i) \right), i=1,....,A \right\} \nonumber \\
&\equiv&
\vec{v}_T^\perp - \left\{ \vec{\dot{v}}_{N,\mathrm{in}}(i)+ \vec{\dot{v}}_{N,\mathrm{out}}(i), i=1,...,A-1 \right\}. 
\end{eqnarray}
The DM particle/nuclear center of mass relative velocity is a c-number,
\be 
\vec{v}_T^\perp ={1 \over 2} \left( \vec{v}_{\chi,\mathrm{in}} + \vec{v}_{\chi,\mathrm{out}}-
\vec{v}_{T,\mathrm{in}}(i)-\vec{v}_{T,\mathrm{out}}(i) \right) \nonumber
\label{eq:cm}
\ee
while the $A-1$ internal nuclear Jacobi velocities $\vec{\dot{v}}_N$ are operators acting on intrinsic nuclear coordinates.
It may be helpful to illustrate this division more explicitly, using one of our interactions, the
axial charge operator $\CO_7$.  We take the simplest example of two nucleons in a nucleus.  Then
\begin{eqnarray}
\vec{v}^\perp \cdot \vec{S}_N &\rightarrow&\sum_{i=1}^2 \frac{1}{2} \left( \vec{v}_{\chi, \rm in} + \vec{v}_{\chi, \rm out} - \vec{v}_{N, \rm in}(i) - \vec{v}_{N, \rm out}(i)\right)  \cdot \vec{S}_N(i) \nonumber \\
 &=& \frac{1}{2} \left( \vec{v}_{\chi, \rm in} + \vec{v}_{\chi, \rm out} - { \vec{v}_{\rm N,in}(1)+\vec{v}_{\rm N,in}(2) \over 2} - { \vec{v}_{N,\rm out}(1) +\vec{v}_{N,\rm out}(2) \over 2} \right) \cdot  \sum_{i=1}^2\vec{S}_N(i) \nonumber \\
  &-&{1 \over 2} \left( {\vec{v}_{N,\rm in}(1)-\vec{v}_{N,\rm in}(2) \over 2}+ 
  {\vec{v}_{N,\rm out}(1)-\vec{v}_{N,\rm out}(2) \over 2} \right) \cdot (\vec{S}_N(1)-\vec{S}_N(2)) \nonumber \\
  &=& \vec{v}_T^\perp \cdot \sum_{i=1}^2\vec{S}_N(i) -\vec{v}_N^\perp \cdot (\vec{S}_N(1)-\vec{S}_N(2) ).
 \end{eqnarray}
yields one term proportional to $\vec{v}_T^\perp$,
\be
\vec{v}_T^\perp \equiv  \frac{1}{2} \left( \vec{v}_{\chi, \rm in} + \vec{v}_{\chi, \rm out} - \vec{v}_{T,\rm in} - \vec{v}_{T,\rm out} \right)  ~~~~\mathrm{where}~~~~\vec{v}_T \equiv {1 \over 2} \sum_{i=1}^2 \vec{v}(i)
\ee
and a second term that depends only on the relative inter-nucleon velocity, and is thus separately
Galilean invariant. This decomposition
can be repeated for A nucleons
\be
\sum_{i=1}^A \frac{1}{2} \left( \vec{v}_{\chi, \rm in} + \vec{v}_{\chi, \rm out} - \vec{v}_{N, \rm in}(i) - \vec{v}_{N, \rm out}(i)\right)  \cdot \vec{S}_N(i)
  = \vec{v}_T^\perp \cdot \sum_{i=1}^A\vec{S}_N(i) - \left[ \sum_{i=1}^A \frac{1}{2} \left( \vec{v}_{N, \rm in}(i) + \vec{v}_{N, \rm out}(i)\right)  \cdot \vec{S}_N(i) \right]_{int} \nonumber
 \ee
where $\vec{v}_T$ is now the target velocity obtained by averaging over $A$ nucleon velocities.
The intrinsic operator on the right can be written in a form that makes the dependence 
on relative nucleon velocities manifest
\be
\label{eq:A1}
 {1 \over 2A} \sum_{i>j=1}^A \left(\vec{S}_N(i)-\vec{S}_N(j)\right) \cdot \left[\left(  \vec{v}_{N, \rm in}(i) + \vec{v}_{N, \rm out}(i)\right)- \left(\vec{v}_{N, \rm in}(j) + \vec{v}_{N, \rm out}(j)\right) \right]
\ee
or, alternatively and trivially, it can be written as the difference of two terms
\be
\label{eq:A2}
\sum_{i=1}^A \frac{1}{2} \left( \vec{v}_{N, \rm in}(i) + \vec{v}_{N, \rm out}(i)\right)  \cdot \vec{S}_N(i) 
-{1 \over 2} \left[ \vec{v}_{T,\mathrm{in}}+\vec{v}_{T,\mathrm{out}} \right] \cdot \sum_{i=1}^A \vec{S}_N(i).
\ee

The above discussion was presented to clarify how the velocity operator is separated into its $\vec{v}_T^\perp$
and intrinsic nuclear pieces, but also to illustrate two assumptions made in our development,
\begin{itemize}
\item The assumption that the WIMP-nuclear interaction is the sum over the individual
WIMP-nucleon interactions leads to two interactions that are separately Galilean invariant,
one constructed from $\vec{v}_T^\perp$ and one constructed from the internal relative nucleon
velocities.  However these two interactions then
 have a common coefficient, $c_7$.  In contrast, if one were to
construct an effective theory at the nuclear level, operators that are separately invariant would
be assigned independent strengths.  It would be interesting to explore whether the
work of \cite{cirigliano,schwenk} on more complicated WIMP-nucleus couplings can be viewed
as adding corrections to the one-body formulation that, in fact, make the two operators independent.
\item While the nuclear matrix elements in the formulas we derive in the text are intrinsic ones,
in fact almost all calculations of the structure of complex nuclei are performed in overcomplete bases
in which the coordinates of all $A$ nucleons appear.  If the underlying single-particle basis is
the harmonic oscillator and if set of included Slater determinants is appropriately
chosen, certain separability properties of the harmonic oscillator allow one to remove the
extra degrees of freedom by numerical means, forcing the center-of-mass into the $1s$ state.
Yet still the basis is expressed in terms of nucleon coordinates.  Largely for this reason, the
intrinsic operator is evaluated using Eq. (\ref{eq:A2}) with the $further$ assumption
that the second, more complicated term in Eq. (\ref{eq:A2}) can be ignored.  This clearly
greatly simplifies the calculation, allowing one to evaluate the nuclear matrix element
from the one-body density matrix.  This kind of approximation -- or more correctly,
simplification -- is used almost universally in nuclear physics, as there is no practical
alternative.  In schematic models it can be shown that the errors induced are typically $o(1/A)$ and
associated with a center-of-mass form factor.  
\end{itemize}
 
 \subsection{Multipole Decomposition}
 In the text leading up to Eq. (\ref{eq:fulldensities}), we formed a WIMP-nucleus interaction by
 assuming the one-body form, as discussed above, interpreting nucleon momenta as
 operators acting on the wave functions of the bound nucleon.  We stressed that the resulting
 interaction has precisely the same form as that conventionally used in spin-independent/spin-dependent (or
 $\CO_1/\CO_4$) analyses, except that a complete set of EFT operators have been included. 
Equation (\ref{eq:fulldensities}), repeated here,
 \begin{eqnarray}
   \sum_{\tau=0,1} \left[
  l_0^\tau~ \sum_{i=1}^A  ~  e^{-i \vec{q} \cdot \vec{x}_i} 
~+~l_0^{A\tau}~ \sum_{i=1}^A ~ {1 \over 2M} \left(-{1 \over i} \overleftarrow{\nabla}_i \cdot  \vec{\sigma}(i) e^{-i \vec{q} \cdot \vec{x}_i}  +  e^{-i \vec{q} \cdot \vec{x}_i}  \vec{\sigma}(i)  \cdot  {1 \over i} \overrightarrow{\nabla}_i \right) \right.  \nonumber \\
 ~+~ \vec{l}_5^\tau \cdot  \sum_{i=1}^A  ~\vec{\sigma}(i)  e^{-i \vec{q} \cdot \vec{x}_i}  
 ~+~ \vec{l}_M^\tau \cdot  \sum_{i=1}^A  ~ {1 \over 2M} \left(-{1 \over i} \overleftarrow{\nabla}_i  e^{-i \vec{q} \cdot \vec{x}_i}  +  e^{-i \vec{q} \cdot \vec{x}_i} {1 \over i} \overrightarrow{\nabla}_i \right)  ~~~~~~~\nonumber \\
~+~ \vec{l}_E^\tau \left. \cdot  \sum_{i=1}^A ~ {1 \over 2M} \left( \overleftarrow{\nabla}_i \times \vec{\sigma}(i)  e^{-i \vec{q} \cdot \vec{x}_i}  +  e^{-i \vec{q} \cdot \vec{x}_i}   \vec{\sigma}(i) \times \overrightarrow{\nabla}_i \right)   \right]_{int} t^\tau(i) ~~~~~~~~~~~~~~~~\nonumber
\end{eqnarray}
where the WIMP tensors appearing above are defined in Eq. (\ref{eq:ls}) and contain all of the 
EFT input in the form of the $c_i$s,
is the starting point for our multipole analysis.   The invariant amplitude is the matrix element of this
interaction
\begin{eqnarray}
\CM_\mathrm{nucleus/EFT} &=& \sum_{\tau=0,1}
  \langle j_\chi, M_\chi; j_N M_N |  \left[
  l_0^\tau~ \sum_{i=1}^A  ~e^{-i \vec{q} \cdot \vec{x}_i}   \right. \nonumber \\
&+& l_0^{A\tau}~ \sum_{i=1}^A ~ {1 \over 2M} \left(-{1 \over i} \overleftarrow{\nabla}_i \cdot  \vec{\sigma}(i)~ e^{-i \vec{q} \cdot \vec{x}_i}  + e^{-i \vec{q} \cdot \vec{x}_i}  \vec{\sigma}(i)  \cdot  {1 \over i} \overrightarrow{\nabla}_i \right)  \nonumber \\
&+&  \vec{l}_5^\tau \cdot  \sum_{i=1}^A  ~\vec{\sigma}(i)~e^{-i \vec{q} \cdot \vec{x}_i} \nonumber \\
&+&  \vec{l}_M^\tau \cdot  \sum_{i=1}^A  ~ {1 \over 2M} \left(-{1 \over i} \overleftarrow{\nabla}_i e^{-i \vec{q} \cdot \vec{x}_i}  +e^{-i \vec{q} \cdot \vec{x}_i} {1 \over i} \overrightarrow{\nabla}_i \right)  \nonumber \\
&+&  \vec{l}_E^\tau \left. \cdot  \sum_{i=1}^A ~ {1 \over 2M} \left( \overleftarrow{\nabla}_i \times \vec{\sigma}(i) e^{-i \vec{q} \cdot \vec{x}_i}  +e^{-i \vec{q} \cdot \vec{x}_i}  \vec{\sigma}(i) \times \overrightarrow{\nabla}_i \right)   \right]_{int} t^\tau(i) ~ | j_\chi, M_\chi; j_N M_N \rangle
\label{eq:IANuc}
\end{eqnarray}
where the subscript $int$ instructs one to take the intrinsic part of the operator (that is, the part depending on the internal Jacobi
coordinates).  

The Hamiltonian can be expressed in terms of nuclear operators carrying good angular momentum and parity and transforming simply under time reversal
by carrying out a standard multipole decomposition.  For the scalar nuclear terms in Eq. (\ref{eq:IANuc}) this involves
the expansion of the plane wave in terms of the Bessel spherical harmonics
\be
M_{JM}(q \vec{x}_i) \equiv j_J(q x_i) Y_{JM}(\Omega_{x_i})
\ee
while for the vector nuclear quantities of the form $\vec{A} e^{i \vec{q} \cdot \vec{x}_i}=\displaystyle\sum_\lambda (-1)^\lambda A_{-\lambda} \hat{e}_\lambda e^{-i \vec{q} \cdot \vec{x}_i}$ one uses Bessel vector spherical harmonics
\be
\vec{M}_{JLM}(q \vec{x}_i) \equiv j_L(qx_i) \vec{Y}_{JLM}(\Omega_{x_i}) \mathrm{~~~~~~~~}
\vec{Y}_{JLM}(\Omega_{x_i}) = \sum_{m~\lambda} Y_{LM}(\Omega_{x_i})~ \vec{e}_\lambda~ \langle L m 1 \lambda|(L1)JM\rangle,
\ee
where $\vec{e}_\lambda$ denotes a spherical unit vector and $A_\lambda = \hat{e}_\lambda \cdot \vec{A}$,
to project out longitudinal, transverse electric, and transverse magnetic components.  After some
algebra $\CM_\mathrm{nucleus/EFT} $ can be written
\begin{eqnarray}
\label{eq:Hfull}
\sum_{\tau=0,1}
  \langle j_\chi, M_{\chi f}; j_N M_{N f} |  \Bigg[  \sum_{J=0}^\infty  \sqrt{4 \pi (2J+1)} (-i)^J  \Big[ l_0^\tau M_{J0;\tau}(q) -i l_0^{A\tau}{q \over m_N} \tilde{\Omega}_{J0;\tau}(q) \Big]  ~~~~~~~~~~~~~~~~~~~~~  \nonumber \\
+\sum_{J=1}^\infty \sqrt{2 \pi (2J+1)} (-i)^J \sum_{\lambda=\pm 1} (-1)^\lambda  \Big[ l_{5 \lambda}^\tau \left( \lambda \Sigma_{J-\lambda;\tau}(q)+ i \Sigma_{J-\lambda;\tau}^\prime(q) \right) ~~~~~~~~~~~~~~~~~~~~~~ \nonumber \\
~~~~~~~~~~~~~~~~~~~~~~~~~~~~~~~~~~~~-i {q \over m_N} l_{M\lambda}^\tau \left(\lambda \Delta_{J-\lambda;\tau}(q)+i \Delta_{J-\lambda;\tau}^\prime(q) \right) ~~~~~~~~~~~~~~~~~~~~~~\nonumber \\
~~~~~~~~~~~~~~~~~~~~~~~~~~~~~~~~~~~~- i {q \over m_N} l_{E\lambda}^\tau \left(\lambda \tilde{\Phi}_{J-\lambda;\tau}(q)+i \tilde{\Phi}_{J-\lambda;\tau}^\prime(q) \right) \Big] ~~~~~~~~~~~~~~~~~~~~~\nonumber \\
+\sum_{J=0}^\infty \sqrt{4 \pi (2J+1)} (-i)^J   \Big[ i l_{5 0}^\tau  \Sigma_{J0;\tau}^{\prime \prime}(q) +{q \over m_N} l_{M0}^\tau \tilde{ \Delta}_{J0;\tau}^{\prime \prime}(q)+ {q \over m_N} l_{E0}^\tau \Phi_{J0;\tau}^{\prime \prime}(q) \Big] \Bigg] | j_\chi, M_{\chi i}; j_N M_{N i} \rangle
\end{eqnarray}
where we have defined the operators as
\begin{equation}
\ O_{JM;\tau}(q)~  \equiv  ~ \sum_{i=1}^A ~ O_{JM}(q\vec{x}_i)~ t^\tau(i)~ .
\end{equation}
The eleven operators appearing above correspond to the charge multipoles of the vector charge (accompanying $l_0$) and axial-vector charge ($l_0^A$) operators,
and the longitudinal, transverse electric, and transverse magnetic projections of the axial-vector spin current (accompanying $\vec{l}_5$), 
vector convection current (accompanying $\vec{l}_M$), and vector spin-velocity current (accompanying $\vec{l}_E$) operators.  As transverse multipoles must carry at least one unit of angular momentum, the multipole sums in those cases begin with $J=1$.   

In elastic transitions the contributing multipoles are severely restricted by the known approximate good parity and CP of nuclear ground states, as detailed in Table \ref{table:symmetries}.  Five of the operators (those not defined in the body of this paper) are eliminated entirely; in other
cases only the even or odd multipoles can satisfy the combined parity and CP requirements.  Thus we
obtain the simpler expression
\begin{eqnarray}
\label{eq:Helastic}
\CM_\mathrm{nucleus/EFT}^\mathrm{elastic} &=& \sum_{\tau=0,1}
  \langle j_\chi, M_{\chi f}; j_N M_{N f} |~ 
\Bigg[ \sum_{J=0,2,...}^\infty  \sqrt{4 \pi (2J+1)} (-i)^J  \left[ l_0^\tau M_{J0;\tau}(q) +  {q \over m_N} l_{E0}^\tau \Phi_{J0;\tau}^{\prime \prime}(q)     \right]   \nonumber \\
&&+\sum_{J=1,3,...}^\infty \sqrt{2 \pi (2J+1)} (-i)^J \sum_{\lambda=\pm 1} (-1)^\lambda  \Big[ i  l_{5 \lambda}^\tau \Sigma_{J-\lambda;\tau}^\prime(q) -i {q \over m_N} l_{M\lambda}^\tau \lambda \Delta_{J-\lambda;\tau}(q) \Big] \nonumber \\
&&+\sum_{J=2,4,...}^\infty \sqrt{2 \pi (2J+1)} (-i)^J \sum_{\lambda=\pm 1} (-1)^\lambda  \Big[ {q \over m_N} l_{E\lambda}^\tau \tilde{\Phi}_{J-\lambda;\tau}^\prime(q) \Big] \nonumber \\
&&+\sum_{J=1,3,...}^\infty \sqrt{4 \pi (2J+1)} (-i)^J   \Big[ i l_{5 0}^\tau  \Sigma_{J0;\tau}^{\prime \prime}(q)  \Big] \Bigg]~| j_\chi, M_{\chi i}; j_N M_{N i} \rangle.
\end{eqnarray}
This expression involves only the six multipole operators of Eq. (\ref{eq:operators}).

The Wigner-Eckart theorem can be used to reduce the nuclear matrix elements.  Then after forming
$|\CM|^2$, averaging over initial nuclear spins, summing over final nuclear spins, and using the 
orthogonality condition imposed by the two three-j symbols obtained in the reduction, one obtains
\allowdisplaybreaks
\begin{eqnarray}
&&{1 \over 2j_N+1} \sum_{M_{N i},M_{Nf}} | \langle j_\chi M_{\chi f}; j_N M_{N f} | ~\CM_\mathrm{nucleus/EFT}^\mathrm{elastic} ~ | j_\chi M_{\chi i}; j_N M_{N i} \rangle |^2 =   {4 \pi \over 2J_i + 1} 
\sum_{ \tau=0,1} \sum_{\tau^\prime = 0,1} \nonumber \\
 &&\left\{ \sum_{J=0,2,...}^\infty  ~ \Bigg(  ~\langle l_0^\tau \rangle \langle l_0^{\tau^\prime } \rangle^* \langle j_N||~ M_{J;\tau} (q)~ || j_N \rangle 
\langle j_N ||~ M_{J;\tau^\prime} (q)~ || j_N \rangle \right.  \nonumber \\
&&~~~+ {\vec{q} \over m_N} \cdot \langle \vec{l}_E^\tau \rangle ~ {\vec{q} \over m_N} \cdot \langle \vec{l}_E^{\tau^\prime} \rangle^*~\langle j_N ||~\Phi^{\prime \prime}_{J; \tau}(q)~ || j_N \rangle  \langle j_N ||~\Phi^{\prime \prime}_{J; \tau^\prime}(q)~ || j_N \rangle \nonumber \\
&&~~~+   {2 \vec{q} \over m_N} \cdot \mathrm{Re} \left[ \langle  \vec{l}_E^\tau \rangle~\langle  l_0^{\tau^\prime} \rangle^* \right]~ \langle j_N ||~ \Phi^{\prime \prime }_{J;\tau}(q)~ || j_N \rangle \langle j_N ||~ M_{J;\tau^\prime} (q) ~|| j_N \rangle   \Bigg) ~~ \nonumber \\
&& +   \sum_{J=2,4,...}^\infty {1 \over 2} \left( {q^2 \over m_N^2}\langle \vec{l}_E^\tau \rangle \cdot
\langle \vec{l}_E^{\tau^\prime}\rangle^* - {\vec{q} \over m_N} \cdot \langle \vec{l}_E^\tau \rangle~ {\vec{q} \over m_N} \cdot \langle \vec{l}_E^{\tau^\prime }\rangle^* \right)    \langle j_N || ~\tilde{\Phi}_{J; \tau}^\prime (q)~ || j_N \rangle   \langle j_N ||  ~ \tilde{\Phi}^{\prime}_{J; \tau^\prime}(q)~ || j_N \rangle \nonumber \\
&& +   \sum_{J=1,3,...}^\infty  \Bigg( \hat{q} \cdot \langle \vec{l}_5^\tau \rangle~ \hat{q} \cdot \langle \vec{l}_5^{\tau^\prime }\rangle^*  ~
\langle j_N ||~ \Sigma^{\prime \prime}_{J; \tau}(q)~ || j_N \rangle
 \langle j_N ||~ \Sigma^{\prime \prime}_{J; \tau^\prime}(q)~ || j_N \rangle  \nonumber \\
 &&~~~+{1 \over 2}  \left( \langle \vec{l}_5^\tau \rangle \cdot \langle \vec{l}_5^{\tau^\prime}\rangle^* - \hat{q} \cdot \langle \vec{l}_5^\tau \rangle ~\hat{q} \cdot \langle \vec{l}_5^{\tau^\prime }\rangle^* \right)  \langle j_N ||~  \Sigma_{J;\tau}^\prime (q)~ || j_N \rangle  \langle j_N || ~\Sigma_{J;\tau^\prime}^{\prime} (q)~ || j_N \rangle \nonumber \\ 
&&~~~+ {1 \over 2} \left( {q^2 \over m_N^2} \langle \vec{l}_M^\tau \rangle \cdot
\langle \vec{l}_M^{\tau^\prime}\rangle^* - {\vec{q} \over m_N} \cdot \langle \vec{l}_M^\tau \rangle ~{\vec{q} \over m_N} \cdot \langle \vec{l}_M^{\tau^\prime }\rangle^* \right)  \langle j_N ||~ \Delta_{J;\tau} (q)~ || j_N \rangle  \langle j_N ||~ \Delta_{J; \tau^\prime}(q)~ || j_N \rangle \nonumber \\
&&  +{\vec{q} \over m_N} \cdot  \mathrm{Re} \left[ i \langle \vec{l}_M^\tau \rangle \times
\langle \vec{l}_5^{\tau^\prime } \rangle^* \right] \langle j_N || ~ \Delta_{J;\tau} (q) ~|| j_N \rangle  \langle j_N || ~\Sigma^{\prime}_{J; \tau^\prime}(q)~ || j_N  \rangle  \Bigg)  \Bigg\}
\end{eqnarray}
where we have used the shorthand for the WIMP matrix elements
\begin{equation}
\langle l  \rangle \equiv \langle j_\chi M_{\chi f} | l | j_\chi M_{\chi i} \rangle
\end{equation}
Note that while our original multipole decomposition was done with a z-axis aligned along $\vec{q}$,
this result is now frame independent as it is expressed entirely in terms of scalar products

Finally, we average over initial WIMP spins and sum over final spins, as in the nuclear case.  The
WIMP tensors involve combinations of 1 and $\vec{S}_\chi$.  As we sum over all magnetic quantum
numbers, the only surviving terms in the bilinear products of the WIMP tensors must transform as spin
scalars, and thus as 1 or as $\vec{S}_\chi^{\, 2}$.  The constant term yields 1.  All cross terms linear
in $\vec{S}_\chi$ must vanish.    The spin terms must be proportional to  $j_\chi (j_\chi+1)$.  The associated
coefficients are easily calculated for the various products
\be
{1 \over 2j_\chi+1} \sum_{m_{\chi_i} m_{\chi_f}} \langle j_\chi m_{\chi_i} | \left\{ \begin{array}{c}
\vec{S}_\chi | j_\chi m_{\chi_f} \rangle \cdot  \langle j_\chi m_{\chi_f} | \vec{S}_\chi \\[0.15cm]
\vec{A} \cdot \vec{S}_\chi | j_\chi m_{\chi_f} \rangle ~~  \langle j_\chi m_{\chi_f} |\vec{B} \cdot \vec{S}_\chi \\[0.15cm]
\vec{A} \times \vec{S}_\chi | j_\chi m_{\chi_f} \rangle \cdot   \langle j_\chi m_{\chi_f} |\vec{B} \times \vec{S}_\chi \\[0.15cm]
\vec{A} \times \vec{S}_\chi | j_\chi m_{\chi_f} \rangle \cdot   \langle j_\chi m_{\chi_f} | \vec{S}_\chi~~~~~~ \end{array} \right\} | j_\chi m_{\chi_i} \rangle =\left\{ \begin{array}{c} 1 \\ [0.19cm] \vec{A} \cdot \vec{B}/3 \\ [0.19cm] 2\vec{A} \cdot \vec{B}/3 \\ 0 \end{array}  \right\} j_\chi(j_\chi+1) 
\ee
The results are further simplified because the resulting scalars $\vec{A} \cdot \vec{B}$ often 
involve longitudinal and transverse quantities or $\vec{q} \cdot \vec{v}_T^\perp$, which vanish.

Executing the associated algebra yields the final result given in Eqs. (\ref{eq:Ham}) and (\ref{eq:HamC}).
The transition probability is expressed as a product of WIMP and nuclear responses functions,
where the former isolates the particle physics in functions that are bilinear in the EFT coefficients,
the $c_i$s.

\begin{table}
\begin{center}
\begin{tabular}{|l|c|c|c|c|}
\hline
\hline
Projection & Charge/current & Operator  & Even J & Odd J  \\
\hline
Charge & Vector charge & $M_{JM}$ & E-E  & O-O \\
Charge & Axial-vector charge &  $\tilde{\Omega}_{JM}$ & O-E & E-O \\
Longitudinal &Spin current & $\Sigma^{\prime \prime}_{JM}$ & O-O & E-E \\
Transverse magnetic &  " & $\Sigma_{JM}$ & E-O & O-E \\
Transverse electric & " & $ \Sigma^{\prime}_{JM}$ & O-O & E-E \\
Longitudinal & Convection current & $\tilde{\Delta}^{\prime \prime}_{JM}$ & E-O & O-E \\
Transverse magnetic &  " & $\Delta_{JM}$ & O-O & E-E \\
Transverse electric & " & $ \Delta^\prime_{JM}$ & E-O & O-E \\
Longitudinal & Spin-velocity current & $\Phi^{\prime \prime}_{JM}$ &E-E & O-O \\
Transverse magnetic & " & $\tilde{\Phi}_{JM}$ & O-E & E-O \\
Transverse electric & " & $\tilde{\Phi}^\prime_{JM}$ & E-E & O-O \\
\hline
\end{tabular}
\end{center}
\caption{The parity-time reversal transformation properties for the eleven operators arising in DM particle scattering
off nuclei.  The nearly exact parity and CP of nuclear ground states restricts the contributing multipoles in elastic
scattering to those that transform under parity and CP as even-even (E-E):  these are the even multipoles of the vector charge operator
$M_{JM}$ and of the longitudinal and transverse electric projections of the spin-velocity current $\Phi^{\prime \prime}_{JM}$
and $\tilde{\Phi}^\prime_{JM}$, and the odd multipoles of the longitudinal and transverse electric projections
of the spin current $\Sigma^{\prime \prime}_{JM}$ and $\Sigma_{JM}^\prime$ and of the transverse magnetic
projection of the convection current $\Delta_{JM}$.}
\label{table:symmetries}
\end{table}

\subsection{Generalizing the Exchange}
Our EFT approach has focused on interactions between the WIMP and nucleus mediated by a
heavy exchange, so that the interaction is pointlike. However, nothing in the treatment of the
WIMP or nuclear vertices depends on this assumption.  We believe the adaptation of this
code for cases in which the exchange is mediated by a photon or other light particle
would be very simple.  This would, of course, require one to add the needed momentum-dependent
propagator to the code.  Once that line is added, however, we see no 
reason that subsequent integrations over phase space
would present any difficulties:  indeed the operator formalism we employ here is the common
formalism for both electron scattering and semi-leptonic weak interactions.  The exchange in the
former is a photon, while the latter is treated as a four-fermion interaction analogous to 
the WIMP case.

\bibliographystyle{h-physrev}
\bibliography{biblio}
\end{document}